\documentclass[galaxies,article,submit,pdftex,moreauthors]{Definitions/mdpi} 

\usepackage[utf8]{inputenc}
\usepackage[english]{babel}
\renewcommand{\linenumbers}{}

\firstpage{1} 
\pubvolume{12}
\issuenum{1}
\articlenumber{1}
\pubyear{2023}
\copyrightyear{2023}
\datereceived{31 October 2023} 
\daterevised{4 December 2023} 
\dateaccepted{13 December 2023} 
\datepublished{25 December 2023} 

\hreflink{https://doi.org/10.3390/galaxies12010001} 
\doinum{10.3390/galaxies12010001}
\pdfoutput=1 

\Title{
Formation of Transitional cE/UCD Galaxies through Massive/Dwarf Disc Galaxy Mergers
}

\TitleCitation{Formation of Transitional cE/UCD Galaxies through Massive/Dwarf Disc Galaxy Mergers}

\Author{{Alexander V. Khoperskov}*,$^{\dagger}$\orcidA{},  {Sergey S. Khrapov} $^{\dagger}$\orcidB{} and {Danila S. Sirotin} $^{\dagger}$\orcidC{}}

\AuthorNames{Alexander V. Khoperskov, Sergey S. Khrapov and Danila S. Sirotin}

\AuthorCitation{Khoperskov, A.V.; Khrapov, S.S.; Sirotin, D.S.}

\address[1]{%
{Volgograd State University,} 
 Universitetsky pr., 100, Volgograd 400062, Russia; {khoperskov@volsu.ru (A.V.K.); khrapov@volsu.ru (S.S.K.); d.sirotin@volsu.ru (D.S.S.) } 
}

\corres{Correspondence: khoperskov@volsu.ru}

\firstnote{These authors contributed equally to this work.} 

\abstract{ 
The dynamics of the merger of a dwarf disc galaxy with a massive spiral galaxy of the Milky Way type have been studied in detail. The remnant of such interaction after numerous crossings of the satellite through the disc of the main galaxy is a compact stellar core, the characteristics of which are close to small compact elliptical galaxies (cEs) or large ultra-compact dwarfs (UCDs). Such transitional cE/UCD objects with an effective radius of 100-200 pc arise as a result of stripping the outer layers of the stellar core during the destruction of a disc dwarf galaxy. Numerical models of the satellite before interaction include baryonic matter (stars and gas) and dark mass.
We use N-body to describe the dynamics of stars and dark matter and Smoothed-Particle Hydrodynamics to model the gas components of both galaxies. The direct method of calculating the gravitational force between all particles provides a qualitative resolution of spatial structures up to 10 pc.
The dwarf galaxy falls onto the gas and stellar discs of the main galaxy almost along a radial trajectory with a large eccentricity. This ensures that the dwarf crosses the disc of the main galaxy at each pericentric approach over a time interval of more than 9 billion years. We vary the gas mass and the initial orbital characteristics of the satellite over a wide range, studying the features of mass loss in the core. The presence of the initial gas component in a dwarf galaxy significantly affects the nature of the formation and evolution of the compact stellar core. Gas-rich satellite gives birth to a more compact elliptical galaxy compared to the merging gas-free dwarf galaxy.
The initial gas content in the satellite also affects the internal rotation in the stripped nucleus.
The simulated cE/UCD galaxies contain very little gas and dark matter at the end of their evolution.
 }


\begin{document}
\noindent 
Citation: {\color{blue} Khoperskov, A.V.; Khrapov, S.S.; Sirotin, D.S. Formation of Transitional cE/UCD Galaxies through Massive/Dwarf Disc Galaxy Mergers // {\bf Galaxies}, 2024, 12 (1), 1. 

\noindent{\url{https://doi.org/10.3390/galaxies12010001}}}


\section{Introduction}

The properties of dwarf galaxies vary over extremely wide ranges of sizes and masses, occupying the entire interval from globular clusters (GCs) to objects with a mass of several billion solar masses and a characteristic radius of several kiloparsecs. The entire zoo of dwarfs is numerous and diverse, being on the one hand the building material in the hierarchical clustering scenario \cite{Battaglia-Nipoti-2022dwarf-DM, Karachentsev-Kaisina-2022Local-Dwarf-Galaxy, Carlsten-etal-2022Local-Dwarf, Karachentsev-etal-2013Catalog, Chilingarian-etal-2007cE-Abell496, Chilingarian-etal-2007dE, McConnachie-2012Dwarf-Galaxies-Local-Group, Henkel-etal-2022Dwarf-Galaxies, Afanasiev-etal-2023KDG64, Makarova-Makarov-2023dwarf}. On the other hand, they themselves are the result of complex evolution, including the mergers and destruction of different objects \cite{Bekki-etal-2001Formation-M32, Deeley-etal-2023formation-cE, Chilingarian-etal-2009cE-Nbody, Newton-etal2023Dwarf-galaxies}.

Galaxy M32 is the prototype for a class of compact elliptical galaxies (cE), which are quite rare \cite{Chilingarian-Bergond-2010cE-UCD, Chilingarian-etal-2009cE-Nbody, Price-etal-2009cE-UCD, Chilingarian-etal-2007cE-Abell496, Chilingarian-Zolotukhin-2015cE-Isolated, Evstigneeva-etal-2007Dwarf-Galaxies-Virgo}. Their formation mechanism is usually attributed to tidal influence, since cE were initially discovered close to giant galaxies in clusters or groups. The typical luminosity of these objects is $\sim 10^9 L_{\odot}$, which is comparable to the luminosity of dwarf elliptical galaxies (dE). However, the small effective radius $r^{(ef\!f)} \sim 200 - 700$\,pc (See, for example, the sample in the work \cite{Chilingarian-etal-2009cE-Nbody}) gives a high surface brightness, which distinguishes these objects from dwarf galaxies (dE). The usual localization of cE galaxies is near central cD-type galaxies. Examples of compact elliptical galaxies near large spiral galaxies are rarer. Let us point out the SBc-type galaxy PGC057129 (See \cite{Bekki-etal-2003UCD-Fornax}), the selection of compact objects near disc galaxies in the work \cite{Norris-etal-2014Bridging-cluster-galaxy} in addition to M32 in M31 \cite{Bekki-etal-2001Formation-M32}.

Ultra-compact dwarf galaxies (UCDs) were discovered at the turn of the 20th and 21st centuries \cite{Minniti-etal-1998open-UCD, Hilker-etal-1999open-UCD, Drinkwater-etal-2000discovery-UCD, Phillipps-etal-2001discovery-UCD, Mieske-etal-2002UCD-GC}. Subsequent studies identify UCDs as intermediate in size and brightness between large globular clusters and small dwarf galaxies with characteristic radii of $10 - 200$ pc and masses of $2\cdot 10^6 - 5\cdot 10^8\,M_\odot$ \cite{Drinkwater-etal-2003compact-dwarf, Zapata-etal-2019merging-clusters, Bruns-Kroupa-2012ultra-compact-dwarf, Zhang-Bell-2017UCD-cE, Mieske-etal-2008nature-UCDs, Paudel-etal-2023UCD-ngc936, Samantha-etal-2014Perseus-Cluster-UCD}.
Moreover, they differ structurally and dynamically both from GCs and from other types of dwarf galaxies \cite{Drinkwater-etal-2003compact-dwarf, Chilingarian-etal-2011UCD-mass}.
UCDs luminosities are between typical values for small dwarf galaxies and large globular clusters.
These objects are brighter than GCs, however, it has been discussed that they may be the extremely bright tail of the globular cluster system's distribution. There is a fairly smooth transition of integral characteristics from the brightest globular clusters to UCDs in the presence of important distinctive morphological properties \cite{Mieske-etal-2002UCD-GC}.

UCDs locations often grouped closer to the center of rich clusters (e.g., the Fornax, Virgo, Coma clusters of galaxies) than dwarf galaxies \cite{Drinkwater-etal-2000discovery-UCD}, which partly indicates their formation mechanism.
The small sizes of cEs and UCDs significantly limit the ability to spatially resolve their internal structure to estimate the effective radius (half-light radius) and mass through the observed stellar velocities dispersion. Therefore, the required resolution is provided only for nearby galaxies \cite{Karachentsev-etal-2004Catalog, Karachentsev-etal-2013Catalog, Carlsten-etal-2022Local-Dwarf, Karachentsev-Kaisina-2022Local-Dwarf-Galaxy}. An interesting case is the late-stage of merging in the disc galaxy NGC 7727 (SABa), where a core with UCD characteristics stands out at a distance of only 480 pc in projection from the center of NGC 7727 with a dynamic mass of $4.2\cdot 10^8\, M_\odot$ \cite{Schweizer-etal-2018UCD-NGC7727}. The orbit of this UCD appears to be highly eccentric.

Various scenarios for the formation of ultra-compact dwarf elliptical galaxies are discussed \cite{Chilingarian-etal-2008UCD, Zapata-etal-2019merging-clusters, Bekki-etal-2003UCD-Fornax, Drinkwater-etal-2003compact-dwarf, Chilingarian-etal-2009cE-Nbody, Mayes-etal-2021UCD-stripping}. We briefly list them below.

\noindent --- Reducing the mass of dwarf elliptical galaxies due to tidal influences may be an important mechanism of galaxy threshing system \cite{Bekki-etal-2003UCD-Fornax, Chilingarian-etal-2009cE-Nbody}. The outer layers of a stellar system can be effectively removed due to tidal forces in such large clusters as the Fornax Cluster, the Coma Cluster, the Perseus Cluster, the Virgo Cluster. The threshing rate is higher for highly eccentric orbits compared to the circular motion of the satellite \cite{Bekki-etal-2003UCD-Fornax}. The compact dwarf core as the basis of the UCD is preserved under strong tidal influences over cosmological times. This mechanism may be sensitive to the initial dark halo profile of the dwarf galaxy \cite{Bekki-etal-2003UCD-Fornax}. UCDs can be born from the core of tidally stripped dwarf elliptical galaxies (dE) or dE/Ns galaxies with low surface brightness \cite{Bekki-etal-2003UCD-Fornax, Chilingarian-etal-2008UCD}.

\noindent ---  
Conditions for UCD formation may be suitable in tidal superclusters that arise from major mergers of galaxies according to the ``Merging Star Cluster Scenario'' \cite{Fellhauer-Kroupa-2005ultramassive-cluster, Kissler-Patig-etal-2006clusters-dwarf, Zapata-etal-2019merging-clusters, Chilingarian-etal-2011UCD-mass}.

\noindent --- The authors of the work \cite{Goodman-Bekki-2018UCD-molecular-clouds} consider the possibility of forming UCDs from supergiant molecular clouds with masses $10^7 - 10^8\, M_\odot$. Hydrodynamic simulations produce clusters with parameters that are typical of ultracompact dwarfs.

\noindent --- Since the sizes of small UCDs and large GCs can be comparable within a sequence with a smooth, continuous distribution of parameters, some common mechanisms for the construction of these objects are discussed, which are the extreme high-luminosity end of the mass function or the result of several associations of large globular clusters \cite{Mieske-etal-2002UCD-GC, Bekki-etal-2002Globular-cluster}.

\noindent ---  The tidal stripping of barred disc galaxy in the cluster cD galaxy can form a compact remnant \cite{Chilingarian-etal-2009cE-Nbody}. This conclusion is confirmed by N-body simulations of the evolution of a large S-galaxy such as the Milky Way, which moves in a gravitational potential with parameters similar to those of the Virgo cluster.

\noindent ---  UCDs radial density profiles satisfy the King model or de Vaucouleurs law \\ ($\propto \exp(-(r/r^{(ef\!f)})^{1/4})$) \cite{Drinkwater-etal-2003compact-dwarf}. Such ultra-compact objects do not have a massive dark component. This property brings UCDs closer to globular clusters \cite{Chilingarian-etal-2011UCD-mass}.

Types cE and UCD appear to form a continuous sequence of galaxies, which allows us to consider transitional cE/UCD objects \cite{Chilingarian-Mamon-2008cE-UCD, Price-etal-2009cE-UCD, Zhang-Bell-2017UCD-cE}. The observed dependence of stellar velocity dispersion as a function of $B$-band absolute magnitudes shows an example of a smooth transition from cE to UCD \cite{Chilingarian-etal-2009cE-Nbody}. The sizes of some recently discovered UCDs are close to $100-200$ pc, which is close to the sizes of small cE galaxies. Examples are NGC936\_UCD as a collapsing satellite of the large disc galaxy NGC 936 with $r^{(ef\!f)}=66.5\pm 17$\,pc \cite{Paudel-etal-2023UCD-ngc936}, the object VUCD7 in the Virgo Cluster with $r^{(ef\!f)} = 96.8$\,pc, $M_* = 8.8\cdot 10^7\, M_\odot$ \cite{Evstigneeva-etal-2007Dwarf-Galaxies-Virgo}, the galaxy NGC0703-AIMSS1 with $r^{(ef\!f)}=165$\,pc, $M_* = 3.1\cdot 10^8\, M_\odot$ \cite{Norris-etal-2014Bridging-cluster-galaxy} etc.

The properties of massive GCs, UCDs, cEs practically fill the entire range of sizes from 5 pc to 600 pc and masses from $2\cdot 10^6 M_\odot$ to $6\cdot 10^9 M_\odot$ \cite{Norris-etal-2014Bridging-cluster-galaxy, Chilingarian-Mamon-2008cE-UCD, Chilingarian-etal-2009cE-Nbody, Zhang-Bell-2017UCD-cE}. The overall sample of massive globular clusters, UCDs and cEs shows a correlation between $r^{(ef\!f)}$ and $M_*$, although these objects lie in a fairly wide area on the plane $(r^{(ef\!f)}, M_*)$ \cite{Zhang-Bell-2017UCD-cE}.
At the same time, objects within the types UCD and cE are divided into two groups according to the mechanism of their formation from different mother galaxies \cite{Norris-etal-2014Bridging-cluster-galaxy}.

The formation mechanism of the recently discovered isolated compact elliptical galaxies may also be based on tidal stripping. After which they were thrown out of groups or clusters due to the dynamic features of the three-body system \cite{Chilingarian-Zolotukhin-2015cE-Isolated, Makarova-etal-2023Isolated-Dwarf}. The authors in \cite{Deeley-etal-2023formation-cE} rely on combining observational data and  IllustrisTNG simulations to analyze the formation of cE and conclude that 32 percent in the sample were formed by the merging of a spiral galaxy near a large galaxy and 68 percent are associated with slow build-up of stellar mass without tidal influence.

{
High-resolution spectroscopy allows you to determine the internal kinematics of UCDs and CEs \cite{Ferre-Mateu-etal-2021UCD-kinemat, Ferre-Mateu-etal-2021UCD-AGN}. Interestingly, the cEs sample in the work \cite{Ferre-Mateu-etal-2021UCD-kinemat} consists of six rotating compact objects, whose the ratio of rotation velocity to velocity dispersion is $0.2-0.5$.
An important achievement is the discovery of black holes with a mass of more than $10^6\,M_\odot$ in the center of ultra-compact dwarf galaxies, which affects the structure and kinematics of the stellar component
\cite{Seth-etal-2014BH-UCD, Ahn-etal-2017BH-UCD, Pechetti-etal-2017cE-BH, Afanasiev-etal-2018BH-UCD, Ahn-etal-2018BH-UCD}.
The presence of supermassive black holes leads to an overestimation of the mass of the stellar population in compact objects \cite{Voggel-etal-2019BH-UCD}. The authors of \cite{Voggel-etal-2019BH-UCD, Ferre-Mateu-etal-2021UCD-AGN} indicate the high prevalence of such systems. Perhaps half of the stripped nuclei contain supermassive black holes \cite{Mayes-etal-2023BH-cE}, the formation of which occurred in the original host galaxy. 
 }

Minor merging in a system with a large disc galaxy is accompanied by the tidal stripping of the dwarf and various disturbances in the main galaxy, the generation of streams, tails, bridges, rings \cite{Vasiliev-etal-2023LMC-MW, Grion-Filho-etal-2021galactic-interaction, Tanaka-etal-2023Interacting-Galaxies, ORyan-etal-2023Catalog-Interacting, Proshina-etal-2022Silchenko}. Stellar and gas streams are an important source of information about the history of galactic evolution and provide verification of various assumptions, including estimates of dark mass parameters \cite{Martínez-Delgado-etal-2023Stellar-Stream, Nibauer-etal-2023Streams}.
Close passages of dwarf objects can significantly influence the kinematics and morphology of the disc components of the main galaxy \cite{Tkachenko-etal-2023GC-bar, Block-etal-2006M32-M31, Katkov-etal-2022rings-S0, Zasov-etal-2017ufn, 
Boselli-etal-2022inner-gas-S0}. There are numerous examples of observed minor mergers in S-galaxies at different cosmological times. Observations highlight cE/UCD near large disc galaxies, for example, NGC936\_UCD \cite{Paudel-etal-2023UCD-ngc936}, NGC~0703-AIMSS1, NGC 0839-AIMSS1, NGC 1316-AIMSS1, NGC 2768-AIMSS1, NGC 3115-AIMSS1, NGC 4350-AIMSS1, NGC 4546-AIMSS1, NGC4594-UCD1 \cite{Norris-etal-2014Bridging-cluster-galaxy}, NGC 0034-S\&S1 \cite{Schweizer-Seitzer-2007Young-Massive-Clusters}, NGC 7252-W3 \cite{Bastian-etal-2013clusters-NGC7252}. Moreover, the object NGC936\_UCD is the compact core of the destroyed dwarf galaxy MATLAS-167 with a stellar tidal flow. There is reasonable evidence of the passage of the M32 through the center of the M31 disc, which may be the cause of some of the observed features in M31 \cite{Block-etal-2006M32-M31}.

Interest in such mergers is fueled by data from the SDSS \cite{Albareti-etal-2017SDSS}, Gaia \cite{Gaia-Collaboration-2016, Gaia-Collaboration-2023}, SEGUE \cite{Yanny-etal-2009SEGUE}, APOGEE \cite{Majewski-etal-2017APOGEE, Imig-etal-2023MW-APOGEE}, the processing of which indicates a possible fall of the satellite Gaia-Sausage-Enceladus (GSE) into our Milky Way about 10 billion years ago \cite{Belokurov-etal-2018stellar-halo, Haywood-etal-2018Stellar-Halo-Gaia-DR2, Evans-etal-2020unusual-Milky-Way, Iorio-Belokurov-2019Gaia, Helmi-etal-2018inner-stellar-halo}. The result of this merger is a significant part of the stellar halo, which has complex spatial and kinematic properties \cite{Lane-etal-2023GSE, García-Bethencourt-etal-2023Milky-Way-simulation, Belokurov-Kravtsov-2022MW, Khoperskov-etal-2023stellar-halo}.
Analysis of the structure of phase space and chemical abundance allows us to identify the history of the different substructures formation due to accretion events. Cosmological simulations and N-body merger simulations allow us to trace the process of changes in these phase features \cite{Khoperskov-etal-2023stellar-halo, Rey-etal-2023VINTERGATAN-GM, Amarante-etal-2022Simulated-Chemodynamical-GSE, Dillamore-etal-2023Stellar-halo-bar, Davies-etal-2023Stellar-halo, Belokurov-etal-2023phase-space-merger, Khoperskov-etal-2023stellar-halo-Chemical}.
The agreement of N-body modeling results with observational data is the basis for determining the parameters of ancient merger \cite{Naidu-etal-2021modeling-GSE}.
There are various estimates of the GSE stellar mass before the start of merging within $(0.15 - 7)\cdot 10^9 M_\odot$ \cite{Limberg-etal-2022Reconstructing-GSE, Lane-etal-2023GSE, Helmi-etal-2018inner-stellar-halo, Belokurov-etal-2018stellar-halo, Vincenzo-etal-2019SGE, Feuillet-etal-2020Gaia–Enceladus–Sausage, Myeong-etal-2019stellar-halo-MW}.

Our efforts in this work are aimed at studying the possibility of the formation of stellar systems with parameters at the junction of small compact elliptical galaxies and large ultra-compact dwarfs (transitional cE/UCD), as a result of the stripping {
of dwarf galaxies } in the field of Milky Way-type disc galaxy.
We study the trajectories of the satellite in the most rigid scenario, when it passes through the disc of the main galaxy at each pericentric approach, which corresponds to the initial orbital eccentricity close to unity. Identifying the role of the initial content of the gas component in the satellite on the formation of transitional cE/UCD is an important part of our research.

\section{Minor merger model}\label{sec:Model}

Minor merger means that the mass ratio of two interacting galaxies satisfies the approximate condition $M^{(Sat)} / M^{(MW)} \le 1/6$, where $M^{(MW)}$ is the mass of a large Milky Way-type galaxy, $M^{(Sat)}$ is the satellite mass inside, for example, optical radius $R_{opt}$). The value of $R_{opt}$ is defined in the model as the radius of a sphere containing approximately 95 percent of the mass of the exponential stellar disc with surface density $\sigma(r) = \sigma_0 \cdot \exp(-r/r_d)$ ($r_d$ is the radial scale of the stellar disc). This profile corresponds approximately to $R_{opt} = 4r_d$.
A key property of a minor merger is the ability to preserve the disc of the main galaxy after the merger is completed.

We use mathematical and corresponding numerical models of the dynamics of multi\-component galaxies, which are described in the works \cite{Khrapov-etal-2021galaxies, Titov-Khoperskov-2022Petersburg, Khrapov-etal-2019Bulletin-Ural, Khoperskov-etal-2021IllustrisTNG}.
Gas movement is based on hydrodynamic equations
\begin{equation}\label{eq:eq-gasdynamics-rho}
 \frac{\partial \varrho}{\partial t} + \vec{\nabla}\left( \varrho \vec{\bf u} \right)  = 0 \,,
\end{equation}
\begin{equation}\label{eq:eq-gasdynamics-u}
 \frac{\partial (\varrho \vec{\bf u}) }{\partial t} + \vec{\nabla}\left( \varrho \vec{\bf u}\otimes\vec{\bf u}\right)  =  - \vec{\nabla} p - \varrho \vec{\nabla}\Psi \,,
\end{equation}
\begin{equation}\label{eq:eq-gasdynamics-E}
 \frac{\partial }{\partial t} \left\{\varrho\, \left( \frac{1}{2} \vec{\bf u}^2 + \varepsilon \right) \right\}
 + \vec{\nabla}\left\{ \varrho \vec{\bf u}\, \left( \frac{1}{2}  \vec{\bf u}^2 + \varepsilon + \Psi \right) \right\}  =  \varrho Q \,,
\end{equation}
where $\varrho$ is the volume gas density, $\vec{\bf u}$ is the velocity vector, $p$ is the pressure, $\varepsilon$ is the internal energy, $\Psi$ is the total gravitational potential from gas, stars and dark matter, $Q$ is the power of the gas heating and cooling by radiation.
The equation of state of an ideal gas with the adiabatic exponent $\gamma = 5/3$ closes the system of equations \eqref{eq:eq-gasdynamics-rho} -- \eqref{eq:eq-gasdynamics-E}.

We use the SPH method to numerically integrate the equations (\ref{eq:eq-gasdynamics-rho}) -- (\ref{eq:eq-gasdynamics-E}) \cite{Khrapov-etal-2021galaxies, Titov-Khoperskov-2022Petersburg, Khrapov-etal-2019Bulletin-Ural, Khoperskov-etal-2021IllustrisTNG}, which has some advantages for modeling interacting multi-component galaxies. Firstly, it is possible to model both the collisional component (gas) and collisionless subsystems (stars and dark matter) in the same way, which is convenient when calculating the gravitational force.
Secondly, there is no need for regularization and setting boundary conditions, in contrast to grid methods of computational fluid dynamics.
Finally, we can easily track the initial origin of gas during the collision of several galaxies, when strong mixing of gas components occurs \cite{Titov-Khoperskov-2022Petersburg}.

Stars and dark matter are considered as collisionless subsystems and are described by the traditional N-body model:
\begin{equation}\label{eq:eq-Nbody}
   \frac{d \vec{\bf v}_i}{dt} =  \vec{\bf f}_{i}  \quad (i=1, \ldots, N) \,,
\end{equation}
where $\displaystyle \vec{\bf v}_i = \frac{d \vec{\bf r}_i}{dt}$ is the velocity of the $i$-th particle, $\vec{\bf r}_i$ is the its radius vector, $\vec{\bf f}_{i}$ is the gravitational force acting on the $i$-th particle from the rest of the mass. 
The interaction between the gas, the stellar component and the dark mass of the satellite is carried out by the method of direct summation of forces between all particles:
\begin{equation}\label{eq:eq-self-gravit}
  \vec{\bf f}_{i} = -\sum\limits_{j=1}^{N} G \, \frac{m_j}{\left(|\vec{\bf r}_{ij}|^2 + \epsilon_c^2\right)^{3/2} }\,\vec{\bf r}_{ij}  \,,
\end{equation}
where $j\ne i$, $m_j$ is the mass of $j$-th particle, $G$ is the gravitational constant, $\vec{\bf r}_{ij} = \vec{\bf r}_{i}-\vec{\bf r}_{j}$,
$\epsilon_c$ is the cutoff radius for the interaction, which ensures collisionlessness for very close passages of two particles, $N$ is the total number of particles in the model, consisting of gas SPH particles ($N^{(gas)}$), star particles ($N^{(star)}$) and dark matter particles ($N^{(DM )}$), $N = N^{(gas)} + N^{(star)} + N^{(DM)}$.  
{
The basic value of the cutoff radius of the gravitational potential in our models is $\epsilon_c = 10$\,pc. }
The direct method of calculating the gravitational interaction of each particle with each (``Particle--Particle'') is very resource-intensive, but gives the best accuracy in modeling a gravitating system, all other things being equal. The use of computing clusters with GPUs for the ``Particle--Particle'' algorithm provides good resolution with the number of particles $2^{20} - 2^{23}$ \cite{Khrapov-Khoperskov-2017SPH-GPUs}.

The value of $Q$ for each gas $i$-th particle in the SPH approach in \eqref{eq:eq-gasdynamics-E} is determined by
\begin{equation}\label{eq:eq-Q}
    Q_i = Q_i^{(ex)} + (1-\alpha_i)Q_i^+ - \alpha_i Q_i^-\,, 
\end{equation}
where $Q_i^{(ex)}$ is the heating of gas particles by external radiation (cosmic rays, stellar radiation),
$\alpha_i$ is the coefficient within $0\le \alpha_i \le 1$,
$Q_i^- = \varrho_i \Lambda(T_i)$ is the cooling function of SPH particles due to radiation, 
$\displaystyle Q_i^+ = \sum\limits_{j=1 (j\ne i)}^{N^{(gas)}} m_j \alpha_j \Lambda(T_j) W(|\vec{\bf r}_{ij}|, h_{ij})$ is the heating of the $i$-th particle by radiation from neighboring particles,
$\Lambda(T_i)$ is the cooling efficiency, depending on the temperature and gas cooling mechanisms \cite{Khoperskov-etal-2013molecular-clouds, Khoperskov-etal-2021heating-cooling},
$W$ is the SPH kernel \cite{Monaghan-Lattanzio-1991Molecular-Clouds}, 
$\displaystyle h_{ij} = \frac{h_i + h_j}{2}$,
$\displaystyle h_i = \frac{4}{3} \left(\frac{m_i}{\varrho_i}\right)^{1/3}$ is the smoothing length of the $i$-th particle.
We do not consider detailed processes of chemical dynamics \cite{Vasiliev-etal-2011Non-equilibrium, Crain-etal-2023Hydrodynamical-Simulations}, limiting ourselves to the energy balance equation (\ref{eq:eq-gasdynamics-E}).

\begin{figure}[!h]
\centering
\includegraphics[width=0.95\hsize]{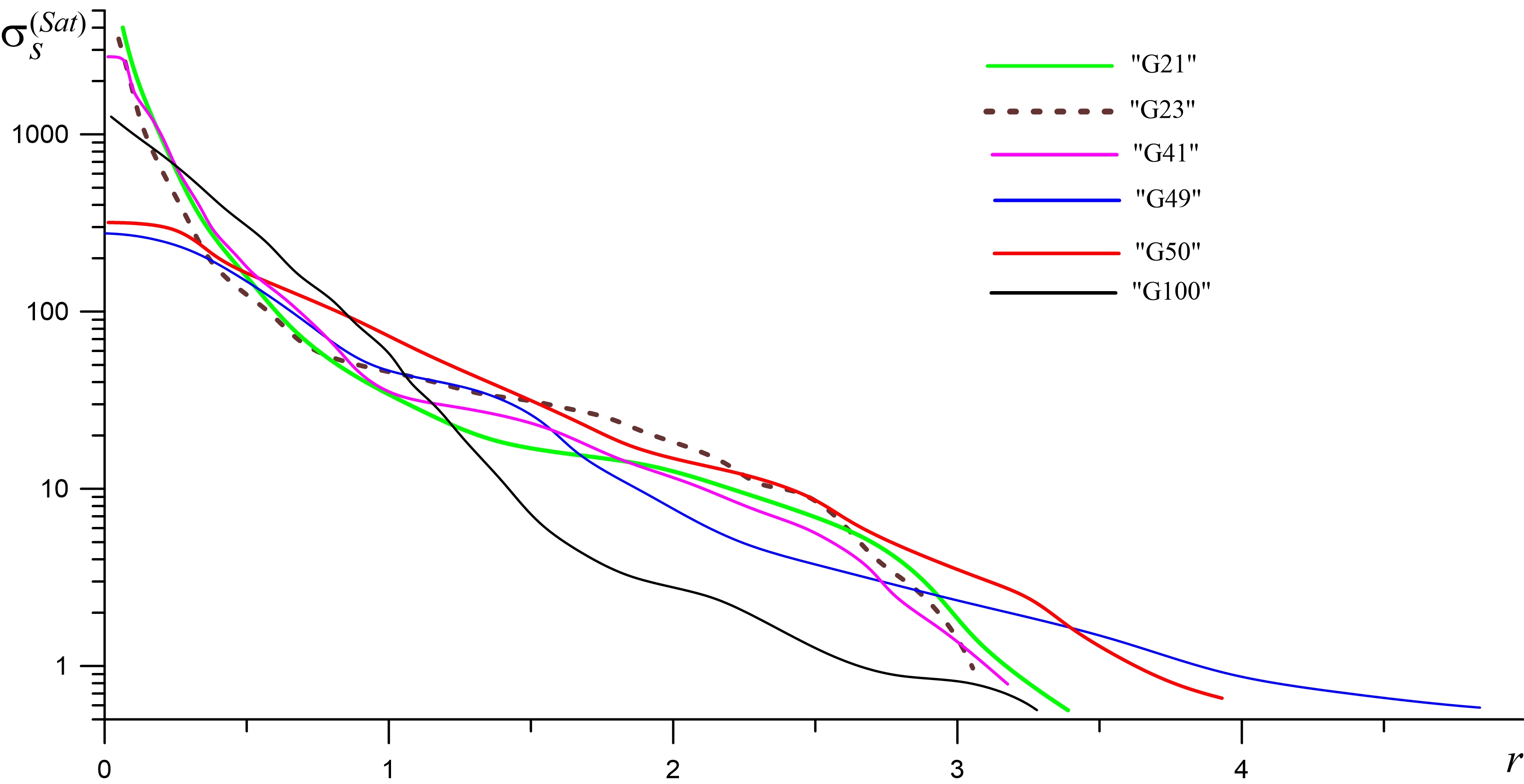} 
\caption{ 
Examples of distributions of azimuthally averaged surface density of stars in different satellite models before the first impact ($[r] = kpc$, $[\sigma_s^{(Sat)}] = M_\odot/pc^2$).
} \label{fig:density-Gaia}
\end{figure}

The two interacting galaxies differ in their mass by a factor of about 20 within their double optical radii ($2R^{(opt)}$). The first main galaxy corresponds to the characteristics of the Milky Way \cite{Khrapov-etal-2021galaxies} and will henceforth be called ``MW model''. We attempt to consider special scenarios for minor merger to MW-type galaxy by tracing the evolution of the stripping of gas-rich dwarf galaxy over more than 9 billion years. Such a model could include a scenario of an event approximately 10 billion years ago associated with the hypothetical object Gaia-Sausage-Enceladus \cite{Belokurov-etal-2018stellar-halo, Deason-2018Stellar-Halo, Haywood-etal-2018Stellar-Halo-Gaia-DR2, García-Bethencourt-etal-2023Milky-Way-simulation, Grand-etal-2020gas-rich-GSE, Helmi-etal-2018inner-stellar-halo, Koppelman-etal-2018Stellar-Halo-Gaia-DR2, Myeong-etal-2018Sausage, Kruijssen-etal-2019MW-GSE}. The parameters of such minor merging seem to be quite typical both in the past and at low redshifts for a wide range of MW type galaxies. The initial physical characteristics of the GSE have not yet been precisely determined, despite significant efforts \cite{Torrealba-etal-2019dwarf-satellite-Gaia, Myeong-etal-2019stellar-halo-MW, Feuillet-etal-2020Gaia–Enceladus–Sausage, Vincenzo-etal-2019SGE, Limberg-etal-2022Reconstructing-GSE, Naidu-etal-2021modeling-GSE, Amarante-etal-2022Simulated-Chemodynamical-GSE, Lane-etal-2023GSE}.
Our satellite model consists of a rotating star-gas disc inside its dark matter halo and is located at a distance of $r_0^{(Sat)}> 100$\,kpc from the center of the main galaxy at the initial time.

\begin{figure}[!h]
\centering
\includegraphics[width=0.49\hsize]{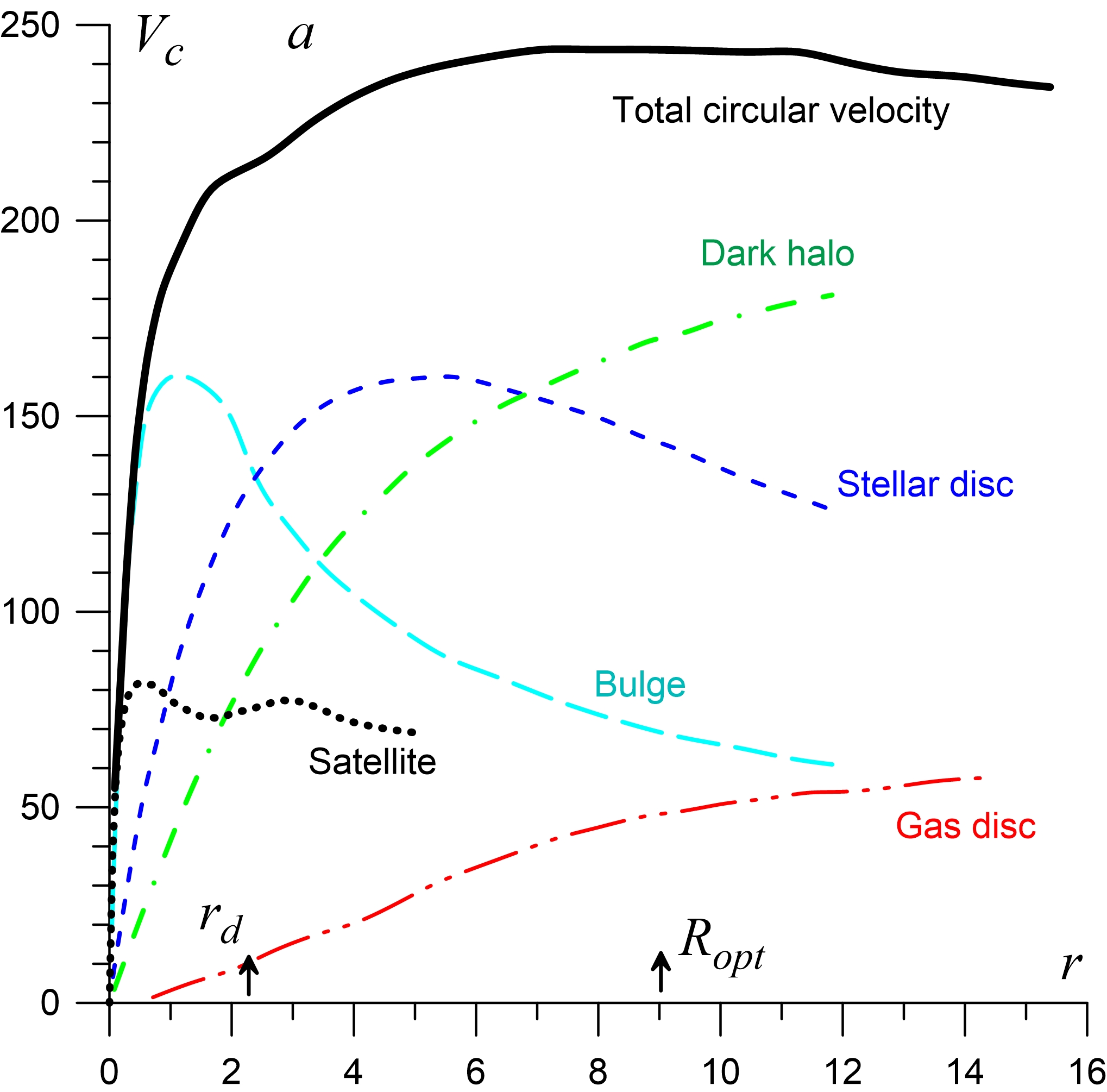} \includegraphics[width=0.49\hsize]{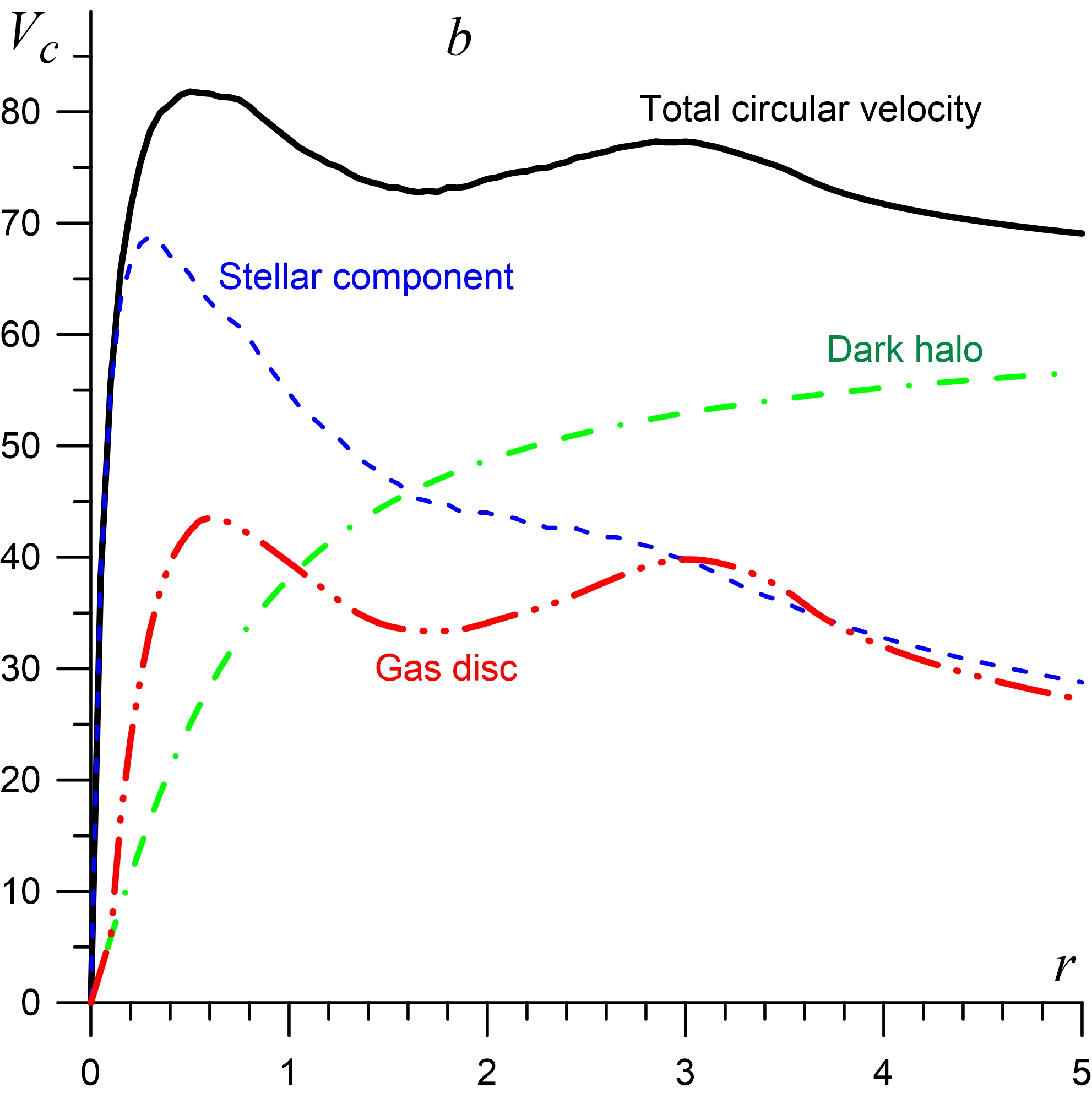} 
\caption{ 
Circular rotation velocities of the MW ($a$), satellite ($b$) models and the corresponding decompositions into galactic components, $[r]=$\,kpc, $[V_c] =$\,km\,s$^{- 1}$.
} \label{fig:Vrot-MW-Gaia}
\end{figure}

{ The initial mass density profiles in the satellite determine the characteristics of the cEs/UCDs. The key role is played by the presence or absence of a nuclear star cluster (NSC) in the satellite \cite{Boecker-etal-2023TNG50-nuclear-star-clusters, Hoyer-etal-2023nuclear-star-clusters} or a compact pseudo bulge. The increased density at the center of the stellar component indicates strong deviations from the exponential profile.
We consider various mass density distributions in the satellite, including both near-exponential discs without NSC and/or pseudo-bulge, and those containing massive nuclear structures with a large concentration of stars. 
 }

{ The initial axisymmetric state of the satellite and MW is constructed based on equilibrium in the radial and vertical directions \cite{Khoperskov-Tyurina-2003MW, Khoperskov-etal-2010AN}. The development of non-axisymmetric gravitational instability in the disc is allowed at a subcritical level. Since the dark halo does not dominate the central region, the stellar bar is formed before the first passage through the host galaxy.
We consider only the centrally symmetric dark halo of the MW, since a triaxial massive halo is capable of generating powerful spiral patterns both in the gas component \cite{Khoperskov-etal-2012halo-gas, Butenko-etal-2022MPCM} and in the stellar disc and redistributing mass along the radius \cite{Khoperskov-etal-2013halo-stellar-disk}. 
The high gas content ($M_g^{(Sat)} \simeq M_s^{(Sat)}$) in some of our models promotes rapid formation of the bar and spiral pattern. The process of the satellite's fall is accompanied by the increasing influence of tidal forces, which also redistributes the matter in the satellite along the radius. Figure \ref{fig:density-Gaia} shows the radial surface density profiles of the stellar component in the companion for some models at time $t>0.4$ billion years before the first crossing of the MW. There are both examples with and without massive concentrated nuclei. All such models contain a stellar bar and a developed spiral structure.  
 }

\begin{table}[h] \caption{Main parameters of the model at the initial time}
    \label{tab:Main-parameters}
    \centering
    \begin{tabular}{l|c}
Parameter  &    \\ \hline
\underline{Milky Way model}: & \\ 
Stellar mass in MW, $M_{s}^{(MW)}$  & $ 3.72 \cdot 10^{10} \,M_\odot$  \\ 
Gas mass, $M_{g}^{(MW)}$ & $0.74 \cdot 10^{10} \,M_\odot$ \\ 
Dark halo mass, $M_{DM}^{(MW)}(r\le 9\, \textrm{kpc}=4r_d^{(MW)})$ & $6.02 \cdot 10^{10} \,M_\odot$ \\
Bulge mass, $M_{b}^{(MW)}$  & $1.0\cdot 10^{10}M_\odot$ \\
Bulge scale, $b^{(MW)}$  & 0.39 kpc \\
Radial scale of stellar disc, $r^{(MW)}_{d_{\,}}$ & $2.25$\,kpc \\  \hline
\underline{Satellite model}: & \\ 
Mass of stars in the satellite, $M_{s}^{(Sat)}$ & $0.093 \cdot 10^{10} \,M_\odot$ \\
Outer boundary of the stellar disc, $R^{(Sat)}_{opt}$  & 2.7 kpc \\
Outer boundary of the gas disc  & 3.5\,kpc \\
Dark halo mass, $M_{h}^{(Sat)}(r\le 5\, \textrm{kpc})$  & $0.372 \cdot 10^{10} \,M_\odot$ \\
Initial azimuthal velocity, $v^{(Sat)}_{0\, rot}$  & $-9.45$\,km$\cdot$sec$^{-1}$  \\
Initial radial velocity, $v^{(Sat)}_{0\, rad}$  & $-9.45$\,km$\cdot$sec$^{-1}$  \\
   \end{tabular}
\end{table}

Figure \ref{fig:Vrot-MW-Gaia} shows the kinematic characteristics of the main galaxy  and the base gas-rich satellite model ``G21''. The circular velocities $V_c(r)$ of each galaxy are depicted by thick lines. The result of the decomposition of galaxy models gives the contributions of various components to $V_c$. We use the MW model from \cite{Khrapov-etal-2021galaxies} (Table \ref{tab:Main-parameters}).
These parameters are fixed in all numerical models.
The surface density of the MW stellar disc with exponential scale $r_d^{(MW)}=2.25$\,kpc gives an optical radius of approximately $R_{opt}^{(MW)}\simeq 4 r_d=9$\,kpc, within which contains more than 90 percent of the stellar disc mass. { The stellar component of the satellite is located before the collision inside $R_{opt}^{(Sat)}\simeq 3$\,kpc. 
 Figure \ref{fig:Vrot-MW-Gaia}\,$b$ contains decomposition of the satellite's circular velocity in the model ``G21'' before its first passage through the host galaxy at time $t = 0.4$ billion years after the start of the simulation. Total circular velocity of the satellite is also added to Figure \ref{fig:Vrot-MW-Gaia}\,$a$ for comparison. 
} 

The gas discs are more extensive, extending up to $2R_{opt}^{(MW)}$ in the case of the MW model.
The satellite's circular velocity is determined primarily by the stellar component within 1.8~kpc (Figure \ref{fig:Vrot-MW-Gaia}$b$). The dark halo dominates only at the periphery of the disc, since we are limited to  dark matter profile with a scale of 0.62 kpc without a cusp, in contrast to the NFW model \cite{Navarro-etal-1996Dark-Halo}. The maximum gas rotation velocity in the satellite is about 80 km\,s$^{-1}$, which is typical for dwarf S-galaxies \cite{Zasov-etal-2021dwarf}. Models of the satellite have different gas contents from complete absence to mass $M_{g}^{(Sat)} = M_{s}^{(Sat)}$. Such gas-rich discs have circular velocities $V_{c\,g}^{(Sat)} \simeq V_{c\,s}^{(Sat)}$ at the periphery of the galaxy, where, however, the halo contribution is dominant (Figure \ref{fig:Vrot-MW-Gaia}$b$). The mass of the Milky Way's dark halo is $M^{(MW)}_h$ inside the optical radius (See Table \ref{tab:Main-parameters}). The parameters of dark matter in both galaxies are constant in all numerical models. 
{The radial volume density profile in the dark halo corresponds to the quasi-isothermal law $\varrho_h(r) = \varrho_{h0}/(1+r^2/a^2)$  with $a=0.62$\,kpc in the satellite and $a = 3 $\,kpc in the MW. }

Our calculations are performed in a Cartesian coordinate system, the center of which is determined by the center of the main galaxy. The initial plane of the MW disc coincides with the plane $z=0$.
We vary the following characteristics of the satellite: the gas mass $M^{(Sat)}_g$, the initial velocity and position of the center of the system, given by the distance to the center MW $r^{(Sat)}_0$ and the angle of incidence $\Theta^{(Sat)}$ (Tables \ref{tab:Main-parameters}, \ref{tab:Gaia-parameters}). The small angle $\Theta^{(Sat)}$ corresponds to incidence from low galactic latitudes.
The angle between the initial planes of the discs for the MW and the satellite is equal to $\beta^{(Sat)}$ ($\beta^{(Sat)} = 0^{\circ}$, if the planes coincide, $\beta^{(Sat)} = 90^{\circ}$ for perpendicular planes).
We also consider models for comparison, in which both disc galaxies are gasless and in which the satellite is spherical. The models also differ in the radial profiles of the components.
Physical characteristics in the work are used both dimensional standard units (pc, $M_\odot$, year) and dimensionless ones, normalized to the following conversion scales: $\ell_M = 3.72\cdot 10^{10} M\ensuremath{_\odot}$, $\ell_r = 9$\,kpc, $\ell_V \simeq 133.7$\,km s$^{-1}$ and $\ell_t \simeq 63.2$\,Myr \cite{Khrapov-etal-2021galaxies}.

\begin{table}[!t]
    \caption{Parameters of some numerical satellite models in addition to the constant characteristics in Table 1}
    \label{tab:Gaia-parameters}
    \centering
\begin{tabular}{l|c|c|c|c|c|c|c}
Name & $M^{(Sat)}_{s_{\,}}$, &  $M^{(Sat)}_{g_{\,}}$, & $r^{(Sat)}_0$, &  $\theta^{(Sat)}$, & $\beta^{(Sat)}$, \\ 
& $10^{10^{\,}}\,M_\odot$\,\,  & $10^{10^{\,}}\,M_\odot$\,\, & kpc  & degrees & degrees \\ \hline
G20 & 0.093 & 0.093 & 100.4 &  70.7 & 0 \\
G21 & 0.093 & 0.093 & 128.5 &  11.3 & 0 \\
G22 & 0.093 & 0 & 100.4 &  70.7 & 0 \\
G23 & 0.093 & 0 & 128.5 &  11.3 & 0 \\
G24 & 0.093 & 0.0465 & 128.5  & 11.3 & 0 \\
G25 & 0.093 & 0.093 & 127.3  & 0 & 0 \\
G27 (Sph) & 0.093 & 0 & 128.5  & 11.3 & -- \\
G28 (Sph) & 0.093 & 0 & 100.4  & 70.7 & -- \\
G30 & 0.093 & 0.093 & 100.4 & 70.7 & 90 \\
G31 & 0.093 & 0.093 & 128.5 &  11.3 & 90 \\
G41 & 0.093 & 0.093 & 128.5 &  11.3 & 0 \\ 
G49 & 0.07 & 0.07 & 128.5 &  11.3 & 0 \\
G50 & 0.085 & 0.085 & 128.5 &  11.3 & 0 
\\
G100 & 0.093 & 0 & 128.5  & 11.3 & 0 
\\
    \end{tabular}
\end{table}

More than 100 computational experiments have been carried out to study minor mergers.
Table \ref{tab:Gaia-parameters} contains the parameters defining the satellite models, which are discussed below. Fixed characteristics are specified in Table \ref{tab:Main-parameters}.

\section{Simulation results}
\subsection{General picture of the minor merging dynamics }\label{subsec:common-picture}

\begin{figure}[!t]
\includegraphics[width=0.95\textwidth]{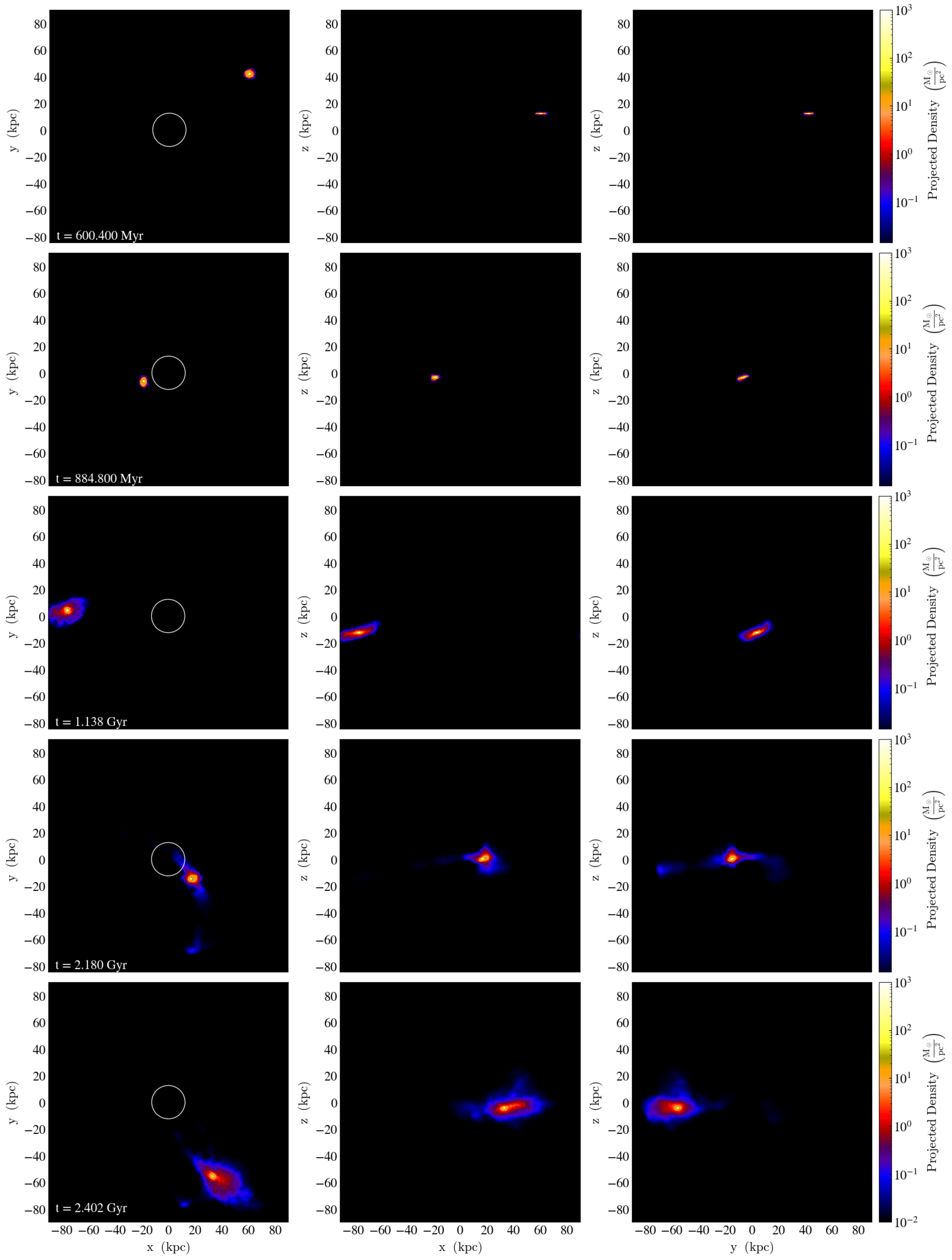}
\caption{
Density distributions of the satellite stars along the line of sight $\Sigma^{(\ell)}_s$ in three projections at five consecutive times ($t=0.60, 0.88, 1.14, 2.18, 2.40$\,billion years) for the ``model G21'' (See Table\,\ref{tab:Gaia-parameters}). The white circle shows the radius where the stellar disc of the main galaxy is located.}
\label{fig:gaia21-star}
\end{figure}

\begin{figure}[!t]
\includegraphics[width=0.95\textwidth]{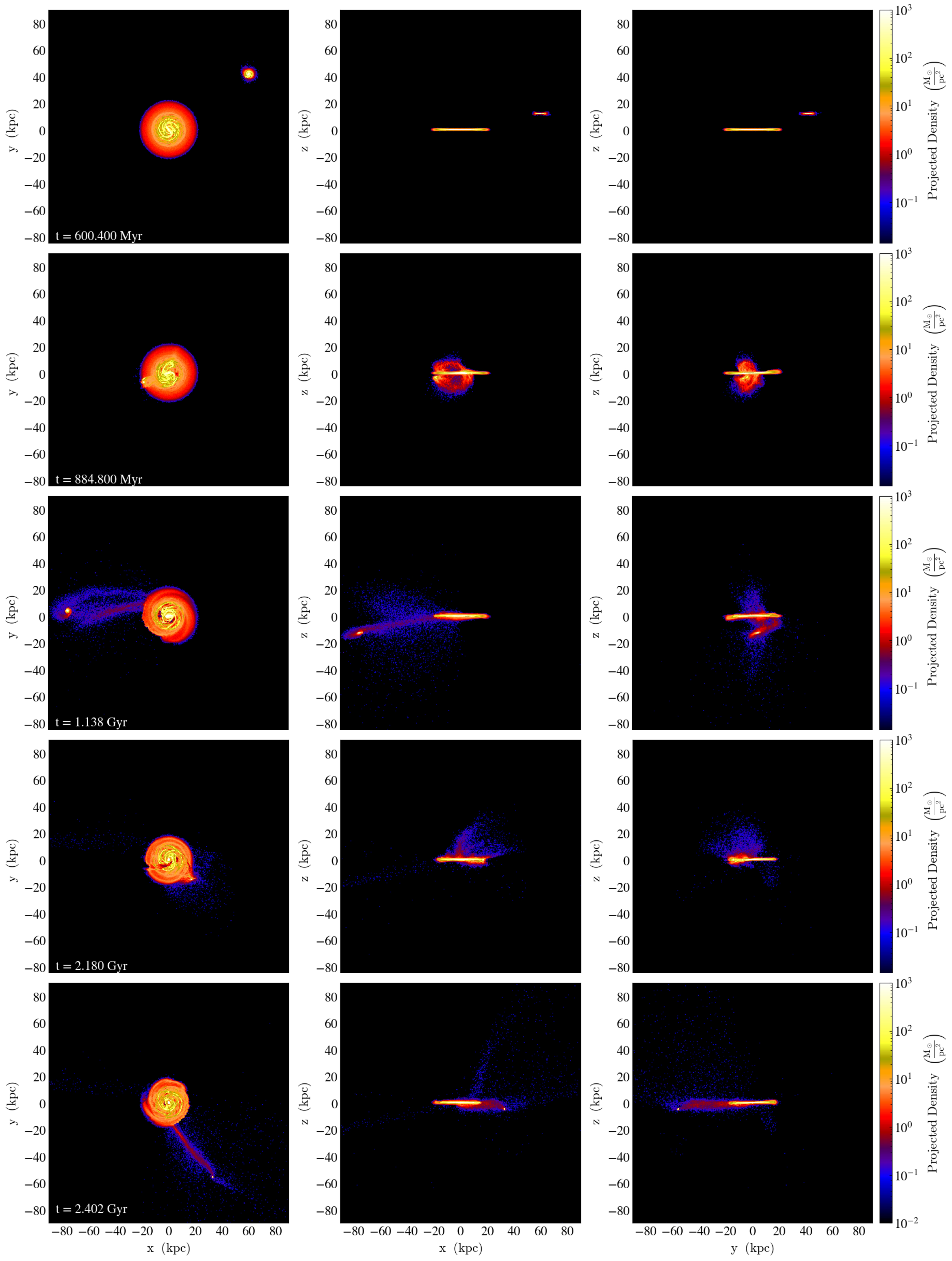}
\caption{
Density distributions of the total gas of two galaxies along the line of sight $\Sigma^{(\ell)}_g$ in three projections at five consecutive times for the model ``G21'', as in Figure\,\ref{fig:gaia21-star}.
}
\label{fig:Gaia21-gas}
\end{figure}

The initial state of the system is chosen so that the satellite passes through the MW disc at the first pericentric approach and each next approach in the numerical experiment. Figure \ref{fig:gaia21-star} shows the characteristic stages of the initial evolution of the stellar component of only the satellite in the process of minor merging at the following time points: 1) pericentric approach of the satellite to the main galaxy ($0.6\cdot 10^9$\,yr); 2) the moment of the first impact ($0.884\cdot 10^9$\,yr); 3) the moment of maximum distance of the satellite from the MW after the first passage through the main disc ($1.138\cdot 10^9$\,yr); 4) moment of the second impact ($2.18\cdot 10^9$\,yr); 5) maximum distance of the satellite’s core from the MW after the second impact ($2.402\cdot 10^9$\,yr). The dwarf galaxy in the experiment ``G21'' falls at a low angle of incidence $\theta^{(Sat)}$. This satellite trajectory disturbs a noticeable part of the MW disc with gas ejected above and below the disc (Figure \ref{fig:Gaia21-gas}).

We do not depict the stars of the main galaxy in Figure \ref{fig:gaia21-star}, which lie within approximately 6 radial scales $r_d^{(MW)}$ for the entire simulation time of 9.5 billion years. This area is highlighted by a white circle with a radius of 12.5 kpc in Figure \ref{fig:gaia21-star}. Only a very small fraction of MW stars are ejected from the original disc, forming a thick disc and part of the stellar halo.
The core of the satellite is clearly visible in yellow with maximum density, surrounded by red and blue areas with lower concentration in Figure \ref{fig:gaia21-star}.
This core is stripped off with each passage through the main disc due to the loss of primarily external particles. As a result, a compact elliptical stellar system begins to form. We will also use the term ``satellite core'' (SC) to refer to proto-cE/UCD at various stages of evolution.

The initial size of the gaseous disc is approximately twice the size of the stellar disc, which enhances the efficiency of the impact interaction between the satellite gas and the host galaxy due to the collisional nature of the gaseous component (Figure \ref{fig:Gaia21-gas}). The satellite loses a noticeable portion of its gas already during the first crossing of the MW disc, throwing gas from the main galaxy out of the disc to a height of up to 10 kpc in both directions. This gas then settles onto the disc. Tidal thin gas bridges and tails are clearly visible in Figure \ref{fig:Gaia21-gas}.

\begin{figure}[!t]
\includegraphics[width=0.95\textwidth]{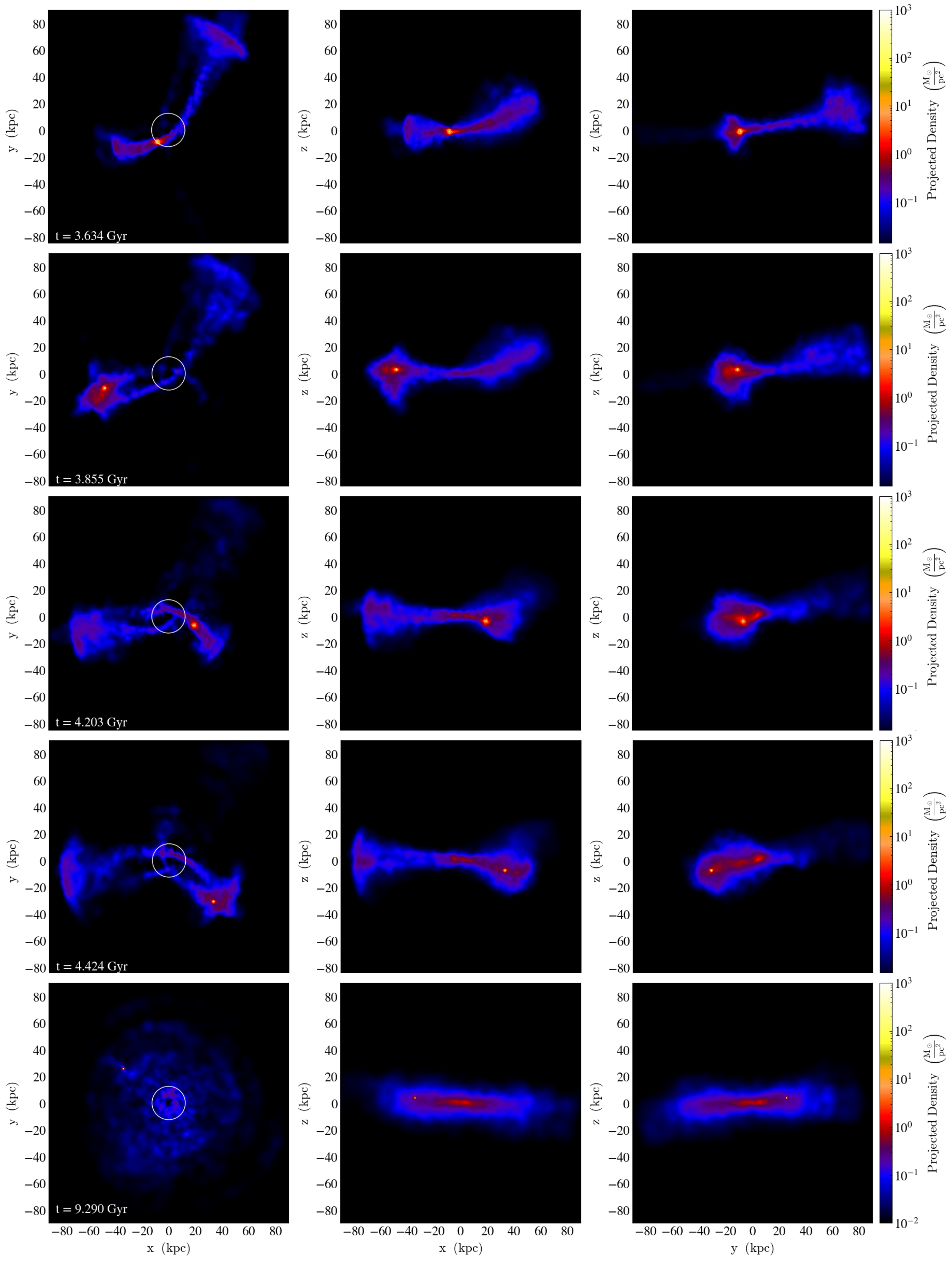 }
\caption{
Density distributions of the satellite stars along the line of sight $\Sigma^{(\ell)}_s$ in three projections as in Figure\,\ref{fig:gaia21-star} at subsequent times (model ``G21'').
}
\label{fig:Gaia21-stars-sp-t6-t10}
\end{figure}

\begin{figure}[!t]
\includegraphics[width=0.95\textwidth]{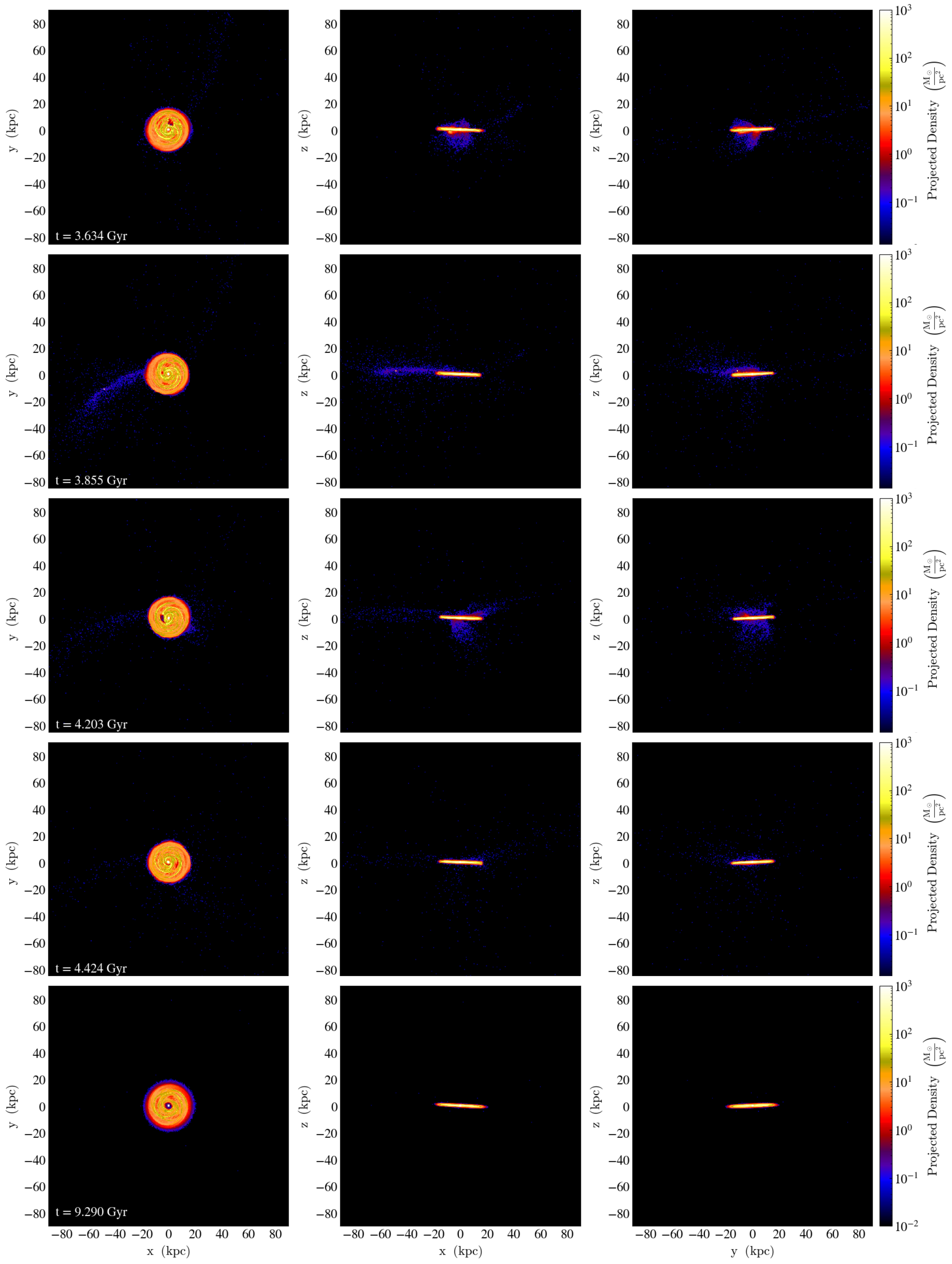}
\caption{
Gas density distributions as in Figure\,\ref{fig:Gaia21-stars-sp-t6-t10} (model ``G21'').
}
\label{fig:Gaia21-gas-sp-t6-t10}
\end{figure}

The further evolution of merging after 2.4 billion years is shown in Figures \ref{fig:Gaia21-stars-sp-t6-t10} and \ref{fig:Gaia21-gas-sp-t6-t10}. Similar three projections of the densities of stars ($\Sigma^{(\ell)}_s$) and gas ($\Sigma^{(\ell)}_g$) are given at five times: 1) fourth crossing of the disc ($3.634\cdot 10^9$\,yr); 2) maximum distance of the satellite core from the MW after the fourth impact ($3.855 \cdot 10^9$\,yr); 3) fifth impact ($4.203\cdot 10^9$\,yr); 4) maximum distance of the core after the fifth passage ($4.424\cdot 10^9$\,yr); 5) the final stage of evolution in the numerical experiment after the 15th passage ($9.29\cdot 10^9$\,yr). We see the continuation of the destruction of the stellar component of the satellite with the formation of a compact dense core. Such a core is clearly visible in the bottom panel of Figure \ref{fig:Gaia21-stars-sp-t6-t10} near the point with coordinates ($x=-34\,{\rm kpc}; y=25.4\,{\rm kpc }; z=3.5\,{\rm kpc}$), which we treat as an object of type UCD/cE.

The stellar halo of minor galaxy stars around the MW disc becomes denser as the dwarf galaxy is stripped away. The oblateness of the inner stellar halo is due to the features of the initial trajectory of the satellite, which invades the main disc from low latitudes, $\Theta^{(GSE)} = 11.3^\circ$ (See Figure \ref{fig:gaia21-star}). The formation of new stellar streams weakens after $4-5$ billion years, as the rate of stars loss to the SC decreases greatly (see below). At later stages, traces of flow structures are clearly visible, which reflects the history of the merger, see, for example, the top view in the plane $(x;y)$ at time $t=9.29$\,Gyr in Figure~\ref{fig:Gaia21-stars-sp-t6-t10}. Various substructures stand out most clearly according to the results of analysis in phase space (See, for example, \cite{Anders-etal-2019gaia, Limberg-etal-2022Reconstructing-GSE, Naidu-etal-2021modeling-GSE, Amarante-etal-2022Simulated-Chemodynamical-GSE, Haywood-etal-2018Stellar-Halo-Gaia-DR2}).

\begin{figure}[!t]
\includegraphics[width=0.99\textwidth]{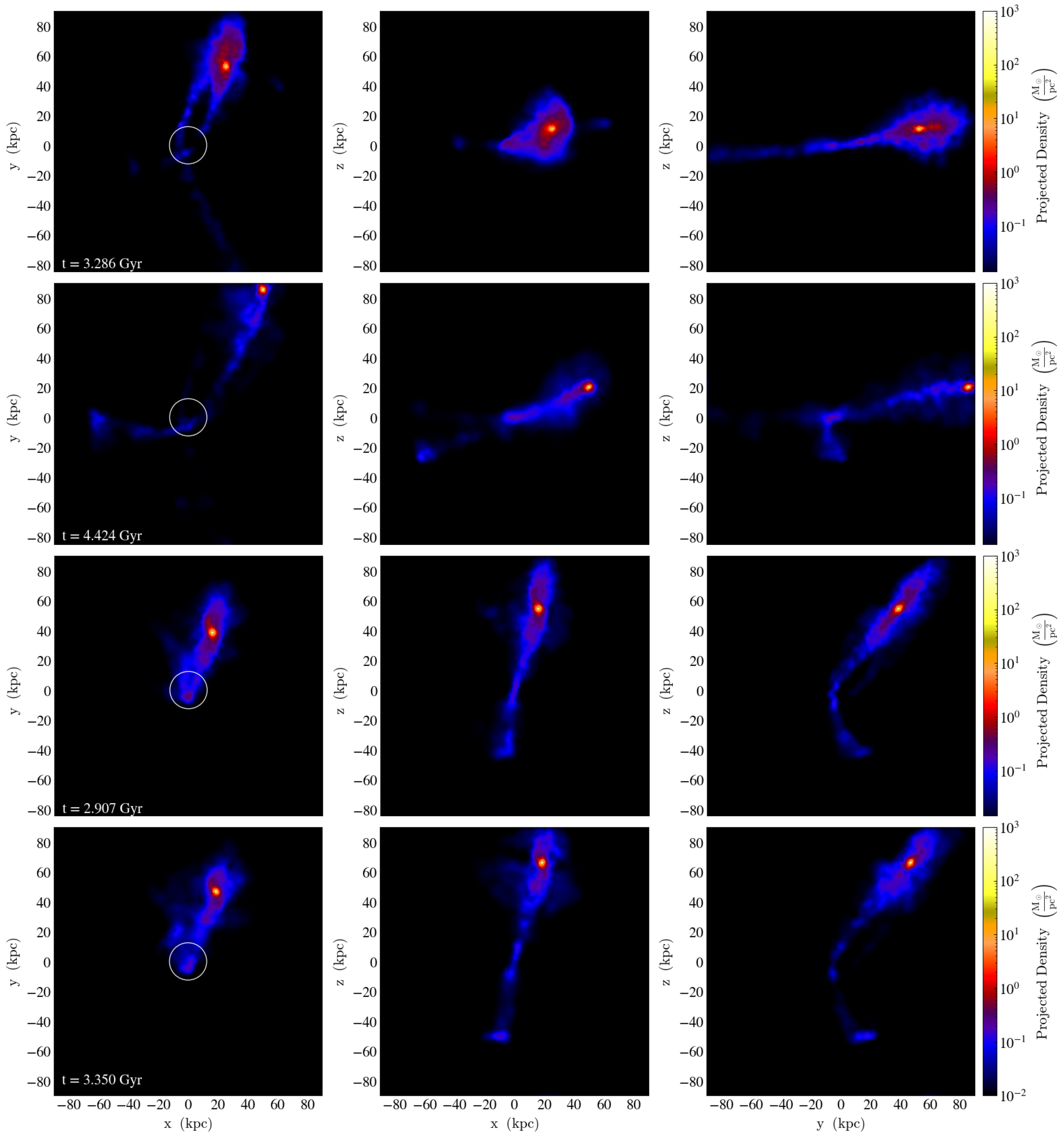}
\caption{
Star density distributions of the satellite $\Sigma^{(\ell)}_s$ after the third  crossing of the MW disc at maximum SC distance for 4 models: 1) ``G21'' (top); 2) ``G23'' (second row); 3) ``G20'' (third row); 4) ``G22'' (bottom).
}
\label{fig:stars-sp-G21-G23-G20-G22-impact3}
\end{figure}

Figure \ref{fig:Gaia21-gas-sp-t6-t10} shows the gas density along the sight line at long times in accordance with the distribution of stars in Figure \ref{fig:Gaia21-stars-sp-t6-t10}.
The density $\Sigma^{(\ell)}_g$ at $t=3.855$ Gyr still highlights the noticeable presence of gas in the SC (compare the yellow spots with coordinates ($x=-53\,{\rm kpc}; y=-10\,{\rm kpc}; z=3\,{\rm kpc}$) on the corresponding panels Figures \ref{fig:Gaia21-stars-sp-t6-t10} and \ref{fig:Gaia21-gas-sp-t6-t10}). At large times, the gas actively leaves the SC, since the density of the gas in the core decreases compared to the gas in the MW and the gravitational well of the satellite core is shallow compared to that created by the MW disc. The gas disc of the main galaxy effectively clears the SC of gas at each crossing.
As a result, the gas component of the satellite is almost completely absorbed by the MW gas disc after 4 Gyr under the conditions of the model ``G21'', which is typical for other models studied. Only a very small fraction of the gas ends up in the intergalactic medium, ranging around 1 percent in different models over long periods of time.

Figure \ref{fig:stars-sp-G21-G23-G20-G22-impact3} shows the distribution of the companion's stellar component in three projections after the third passage through the disc of the main galaxy. These images demonstrate the influence of model parameters on merger dynamics. The two basic models ``G20'' and ``G21'' initially contain a lot of gas and differ only in their initially fall trajectories at close orbital eccentricity values. The angle of attack in the first model is close to normal ($\theta^{(GSE)}\simeq 71^\circ$), and the second model has $\theta^{(GSE)}\simeq 11^\circ$ (See Table \ref{tab:Gaia-parameters}). We compare these two models with the corresponding calculations without gas (models ``G22'' and ``G23''). These pictures show the influence of the gas in the satellite on the dynamics of the merger. Pair of models ``G21'' and ``G23'' provide a noticeable braking effect from the gas, since the core in the ``G23''  goes to a maximum distance from the MW center of 101 kpc instead of 59 kpc for the ``G21'' (Compare the first and second rows in Figure \ref{fig:stars-sp-G21-G23-G20-G22-impact3}).
The similar difference between the pair of simulations ``G20'' and ``G22'' is 68 kpc and 83 kpc, respectively.

\subsection{Formation and characteristics of transitional cE/UCD}\label{subsec:massa-cE}

We will consider in more detail the properties of the satellite's cores, which stand out in the Figures \ref{fig:gaia21-star} -- \ref{fig:stars-sp-G21-G23-G20-G22-impact3} in the form of small yellow spots. Their masses and sizes fall within the range between small compact elliptical galaxies (cEs) and the largest ultra-compact dwarf galaxies (UCDs), which are classified as the transitional UCD/cE type.

\begin{figure}[t]
\includegraphics[width=0.97\hsize]{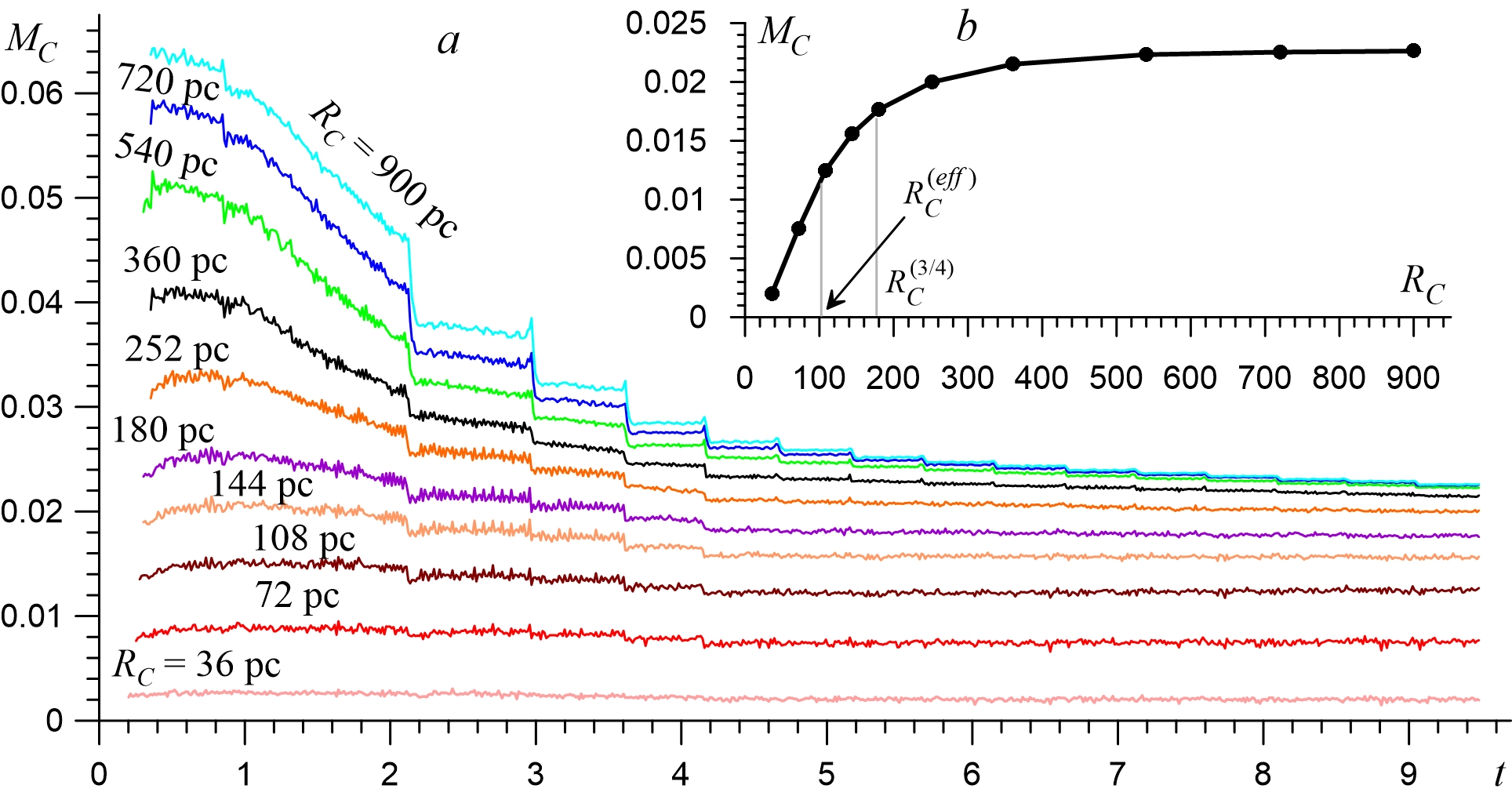}
\caption{
Mass of stars in the SC ($[M_C] = 10^{10}\,M_\odot$) inside a fixed radius ($[R_C]=$\,pc) vs. time in the model ``G21'' with high initial gas content. $b$ --- Mass of stars inside the sphere with radius $R_C$ for the final state ($t>9\cdot 10^9$\,years).
\label{fig:mass-core-radius} }
\end{figure}

\begin{figure}[t]
\includegraphics[width=0.97\hsize]{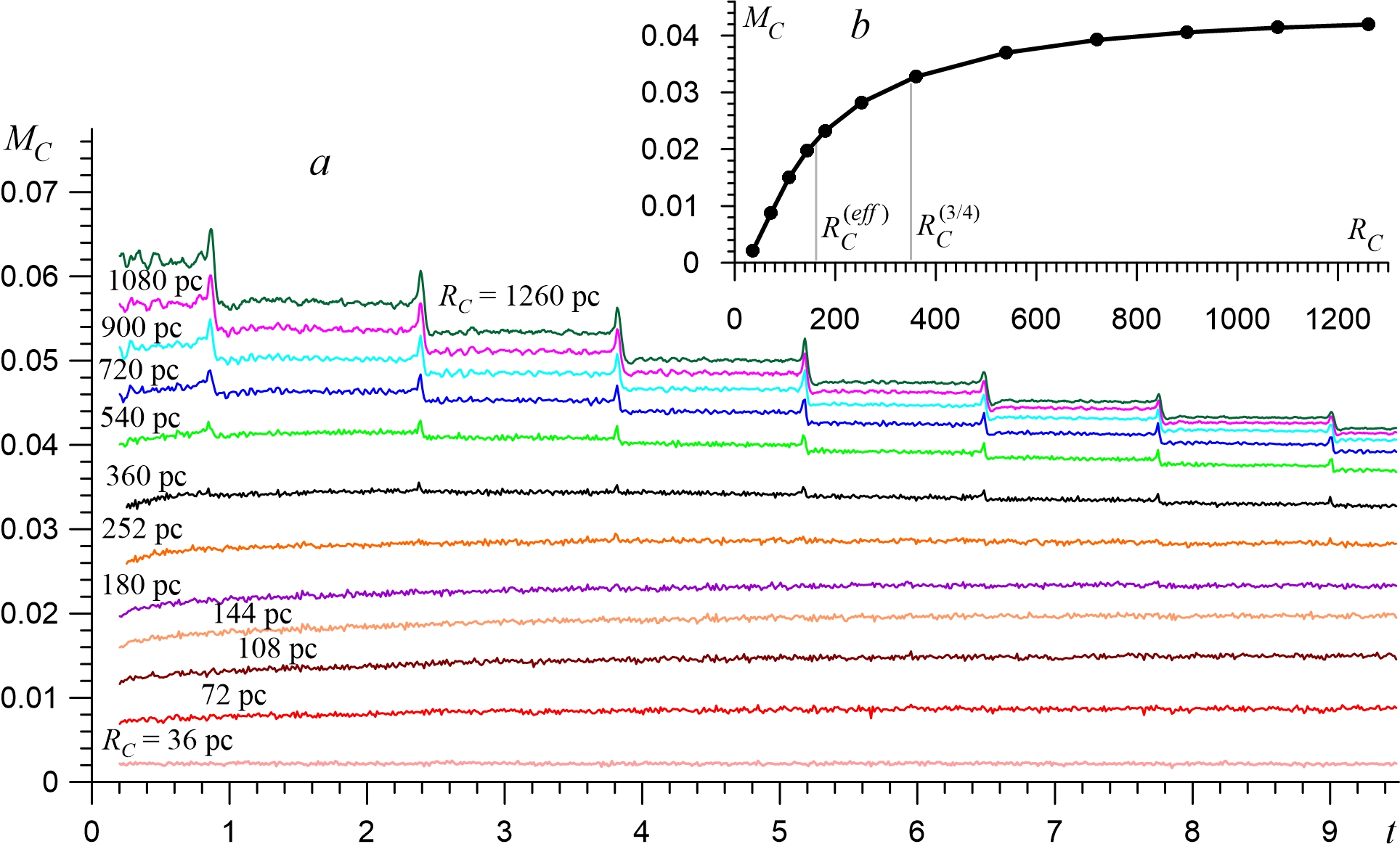}
\caption{As in Figure \ref{fig:mass-core-radius} for the model ``G23'' without gas.
\label{fig:Ms_core_r-Gaia23}
}
\end{figure}

The first problem is related to isolating the size of the satellite's core in a numerical model throughout its evolution. The SC center is determined by the density maximum at some point $\vec{\bf r}_C$, as seen in Figures \ref{fig:gaia21-star} -- \ref{fig:Gaia21-gas-sp-t6-t10}.
We calculate the mass inside spheres centered at $\vec{\bf r}_C$ with different radii $R_C$ as a function of time. Such time dependences of the mass $M_C(t;R_C)$ are shown in Figure \ref{fig:mass-core-radius}, which are typical for all models. Passages of the SC through the main disc are clearly marked by jumps in the dependences $M_C(t)$, especially for large radii $R_C$ (See Figure \ref{fig:mass-core-radius}). The lost mass in the core is higher at the first crossing and decreases with time. The determination of the effective core radius $R_{C}^{(ef\!f)}$ contains some uncertainty due to differences when using either the bulk density of matter in the core ($\varrho_C(r)$) or the surface density along the sight line ($\Sigma^{(\ell)}_s$). We assume below that within the radius $R_{C}^{(ef\!f)}$ is half the mass based on bulk density calculations.

Figure \ref{fig:Ms_core_r-Gaia23} is analogous to Figure \ref{fig:mass-core-radius} for the model ``G23'' in the initial absence of gas in the satellite. The efficiency of core mass loss is reduced by approximately 2 times. As a result, the final size $R_C^{(ef\!f)}$ (or $R_C^{(3/4)}$) and the corresponding core mass $M_C$ is approximately 2 times larger. Narrow peaks on the functions $M_C(t)$ at the moments of passage of the core through the disc are determined by the geometric factor, since the stellar mass of the main galaxy also falls inside the fixed sphere $R_C$. The time such particles velocity inside the core is short and they quickly leave the radius $R_C$, since there is no gravitational connection due to very different kinematic characteristics (See, for example, the curve for $R_C = 1260$ pc in Figure \ref{fig:Ms_core_r-Gaia23}).

 The considered minor merger leads to the formation of a compact/ultra-compact elliptical-type star system from a dwarf spiral galaxy.
Simulation of the interaction of a satellite without gas gives a more massive and larger core for a given initial mass of the stellar component $M_s^{(Sat)}$. The presence of gas can create a more compact core.
Approximately $30-50$ percent of the satellite's stellar mass goes towards building the elliptical core. { The rest of the matter scatters, forming primarily a stellar halo around the main galaxy. }

Figure \ref{fig:mass-core-radius} gives a visual representation of the mechanism of core stripping due to interaction, when mass loss occurs due to the outer layers of the core. The mass decreases most dramatically during approximately the first 4 billion years within the sphere of 300–900 ps. Then the SC mass almost stops changing, so the effective radius is approximately $R_{C}^{(ef\!f)}\simeq 100$ pc and 3/4 of the mass is inside $R_{C}^{(3/4 )} \simeq 180$ pc. The dependence in Figure \ref{fig:mass-core-radius}$b$ shows the mass content $M_C$ inside a sphere of radius $R_C$ in the interval $t=9-9.5$ billion years, highlighting the radii $R_{C}^{ (ef\!f)}$ and $R_{C}^{(3/4)}$.

\begin{figure}[h]
\includegraphics[width=0.42\hsize]{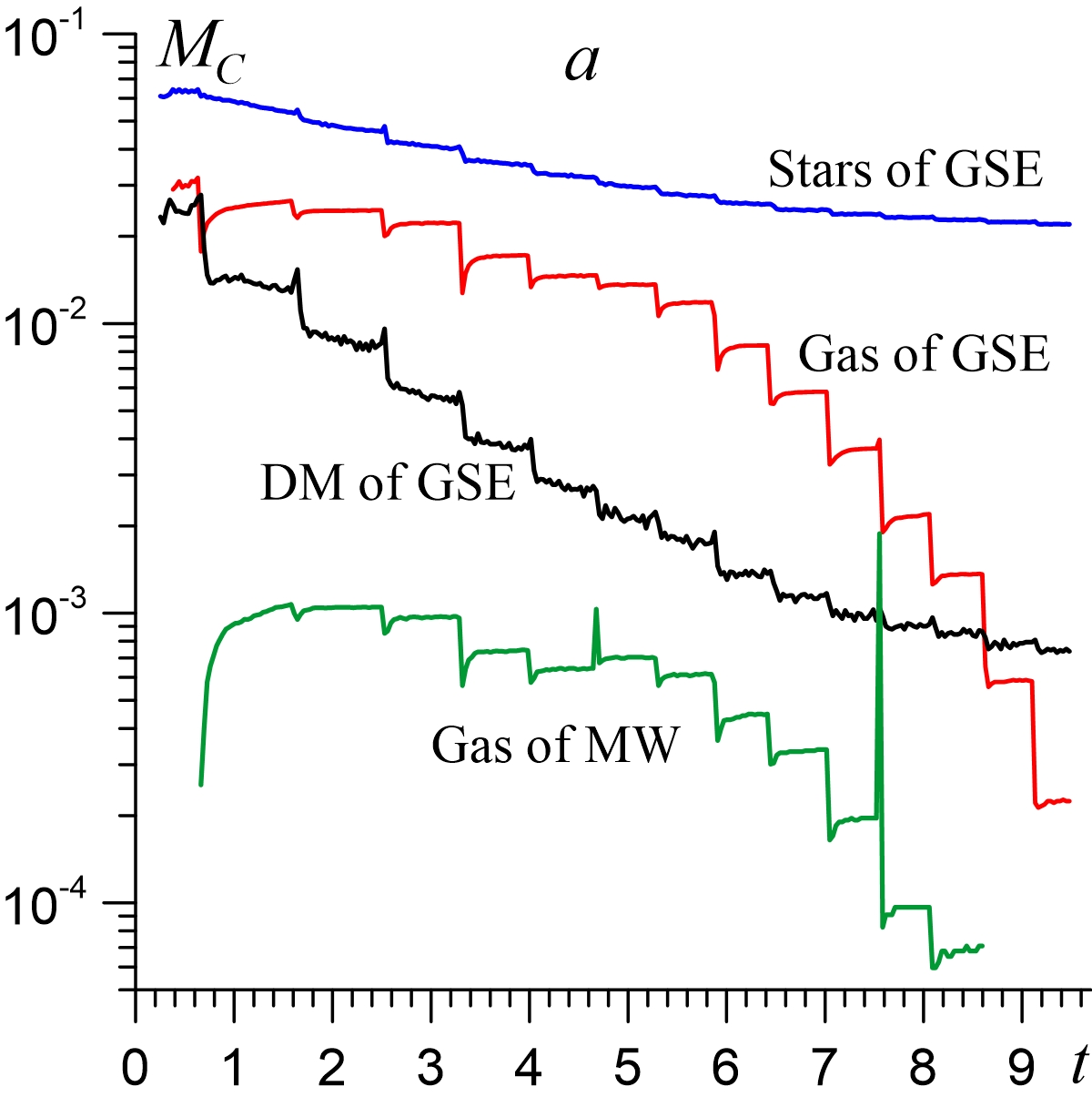} \hskip 0.05\hsize
\includegraphics[width=0.42\hsize]{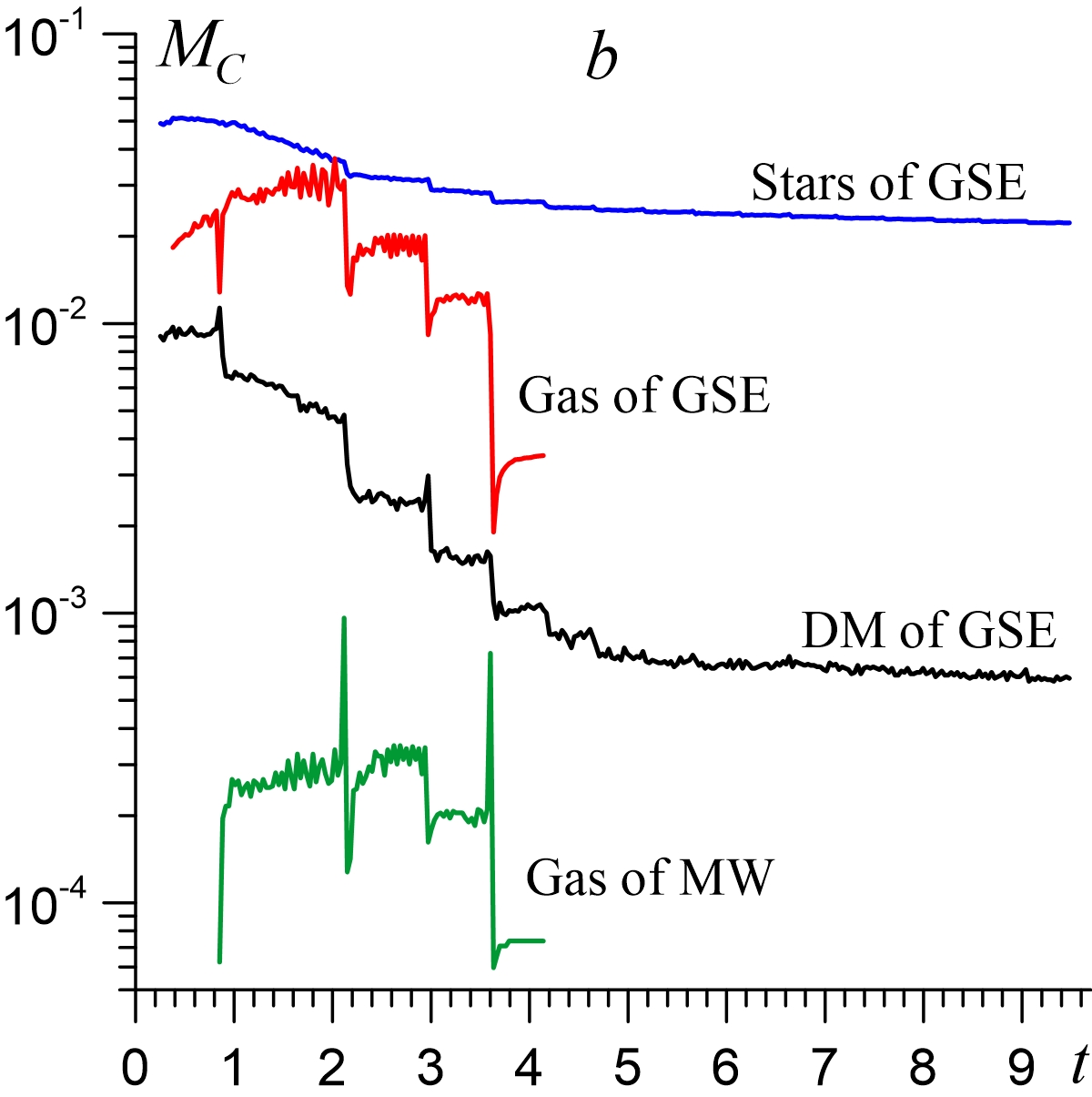} 
\caption{
Masses of various components of the forming SC as a function of time in the models ``G20'' with $R_C = 900$\,pc ($a$) and ``G21'' with $R_C = 540$\,pc ($b$).
} \label{fig:Msgdm_core_r-Gaia20-21}
\end{figure}

The satellite contains three components (stars, gas and dark mass), which react differently to the process of interaction with the main galaxy over 10 billion years (Figure \ref{fig:Msgdm_core_r-Gaia20-21}).
The times of crossing of the disc are well recorded by characteristic jumps of decreasing mass inside a sphere with a given radius. The satellite's gas is retained within the core after three passes through the MW disc in the model ``G21'' (See the red line in Figure \ref{fig:Msgdm_core_r-Gaia20-21}$b$). The fourth crossing sweeps away gas almost completely from the core. The fall of the satellite from high galactic latitudes in the model ``G20'' can preserve gas in the core for longer times (Figure \ref{fig:Msgdm_core_r-Gaia20-21}$a$). There is an effect of a secondary increase in the mass of gas in the core after passing through the main disc due to the capture of matter at high galactic altitudes. The mass of gas in the core (red curves in Figure \ref{fig:Msgdm_core_r-Gaia20-21}) sharply decreases when the core collides with the disc and then quickly increases to an approximately constant level until the next interaction event. Real capture of gas from the main galaxy by the core is possible with different merging parameters (See green lines in Figure \ref{fig:Msgdm_core_r-Gaia20-21}). However, the mass of MW gas in the core of the satellite is very small at all stages of evolution. The narrow upward peaks in the dark mass are a consequence of the geometric method of calculating $M_C$, as in the case of the stellar component. It must be emphasized that the masses in Figure \ref{fig:Msgdm_core_r-Gaia20-21} are calculated inside spheres of large radius, where almost all of the gravitationally bound mass of the core is guaranteed to be located. The effective radii $R_{C}^{(ef\!f)}$ are noticeably smaller than the considered values (See Figures \ref{fig:mass-core-radius}, \ref{fig:Ms_core_r-Gaia23}).

The dark mass does not have a significant effect on the dynamics of SC formation (See the black curves in Figure \ref{fig:Msgdm_core_r-Gaia20-21}) and as a result, the formed compact star system contains a few percent of the dark mass.
The black solid line shows the DM within 900 pc and the dotted line is calculated for a sphere with a radius of 5 kpc, which determines the region of the initial location of dark matter in the satellite.
As we can see, the rate of DM loss is maximum compared to stars and gas.

\begin{figure}[h]
\includegraphics[width=0.9\hsize]
{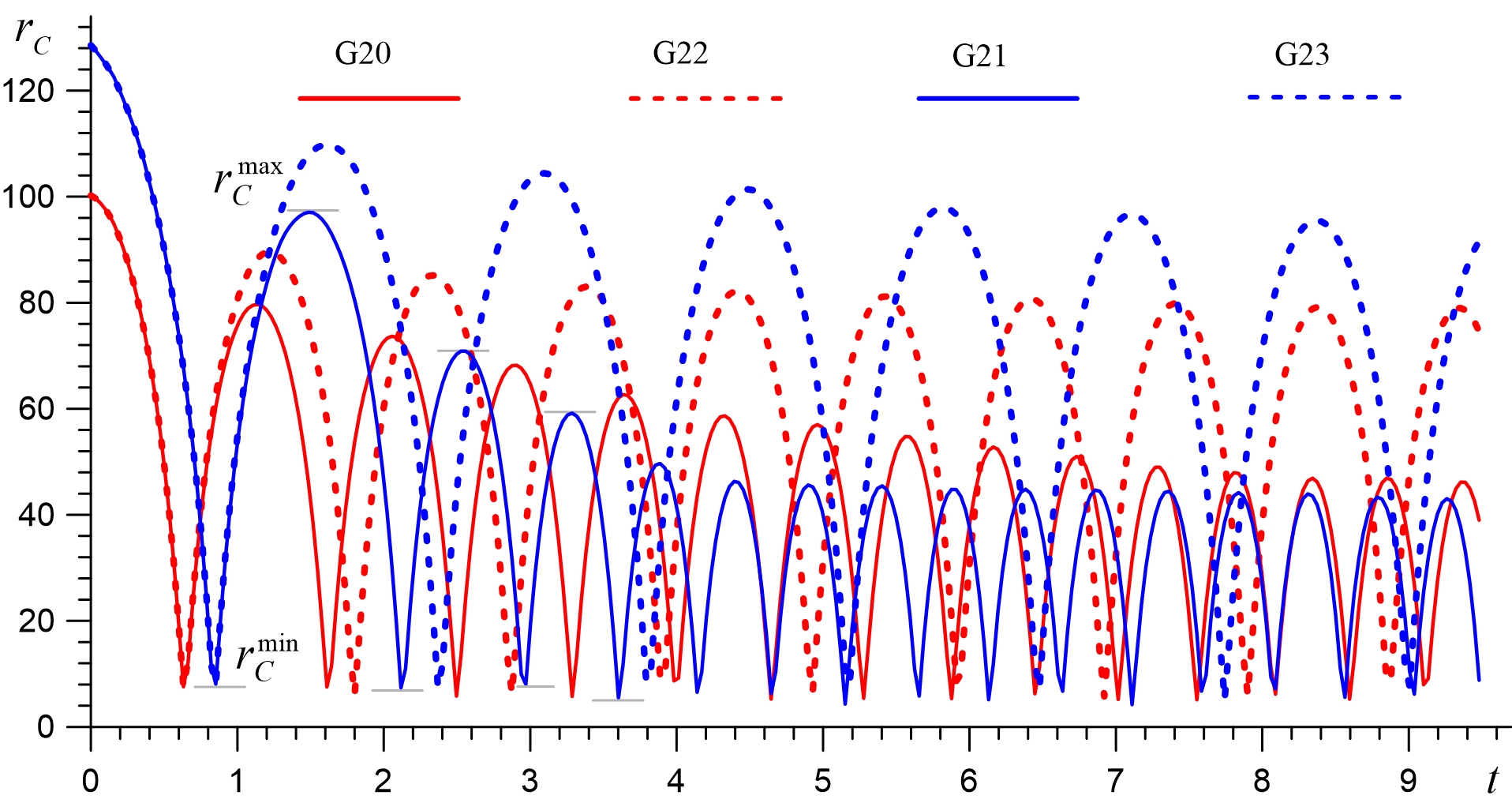} 
\caption{
Dynamics of distances between the SC and MW centers, $[r_C] = {\rm kpc}$, $[\,t\,] = $\,billion years.
} \label{fig:R_core_20-21-22-23}
\end{figure}

The trajectories of the core are complex due to the heterogeneity and asymmetry of the gravitational potential. The distance between the centers MW and SC is equal to $r_C$. The function $r_C(t)$ is characterized by a maximum distance ($r_C^{(\max)}$, apocenter) and a minimum value ($r_C^{(\min)}$, pericentric approach), indicating the radius in the MW disc at which the core crosses through the disc (Figure \ref{fig:R_core_20-21-22-23}). The dependences in Figures \ref{fig:Msgdm_core_r-Gaia20-21} and \ref{fig:R_core_20-21-22-23} clearly distinguish two stages of evolution of the distance $r_C(t)$ in models with a large gas content (models `` G20'' and ``G21''). There is a rapid decrease in local maxima initially in the presence of gas. The decrease in $r_C^{(\max)}$ slows down significantly after the fourth MW crossing. The role of gas is shown in Figure \ref{fig:R_core_20-21-22-23}, where the dependences $r_C(t)$ are compared in pairs for the models ``G20''\,--\,``G22'' and ``G21'  ''\,--\,``G23''. The difference in gas content in the satellite models leads to a strong change in the trajectory due to the sweeping of gas from the SC. Models without gas (dashed lines) give small changes in the maximum core distance $r_C^{(\max)}$  from the MW center over 9 Gyr. The gas-rich satellite (solid lines) moves in a more limited region and $r_C^{(\max)}$ is 50 -- 60 percent smaller compared to models without gas.

\begin{figure}[h]
\includegraphics[width=0.48\hsize]{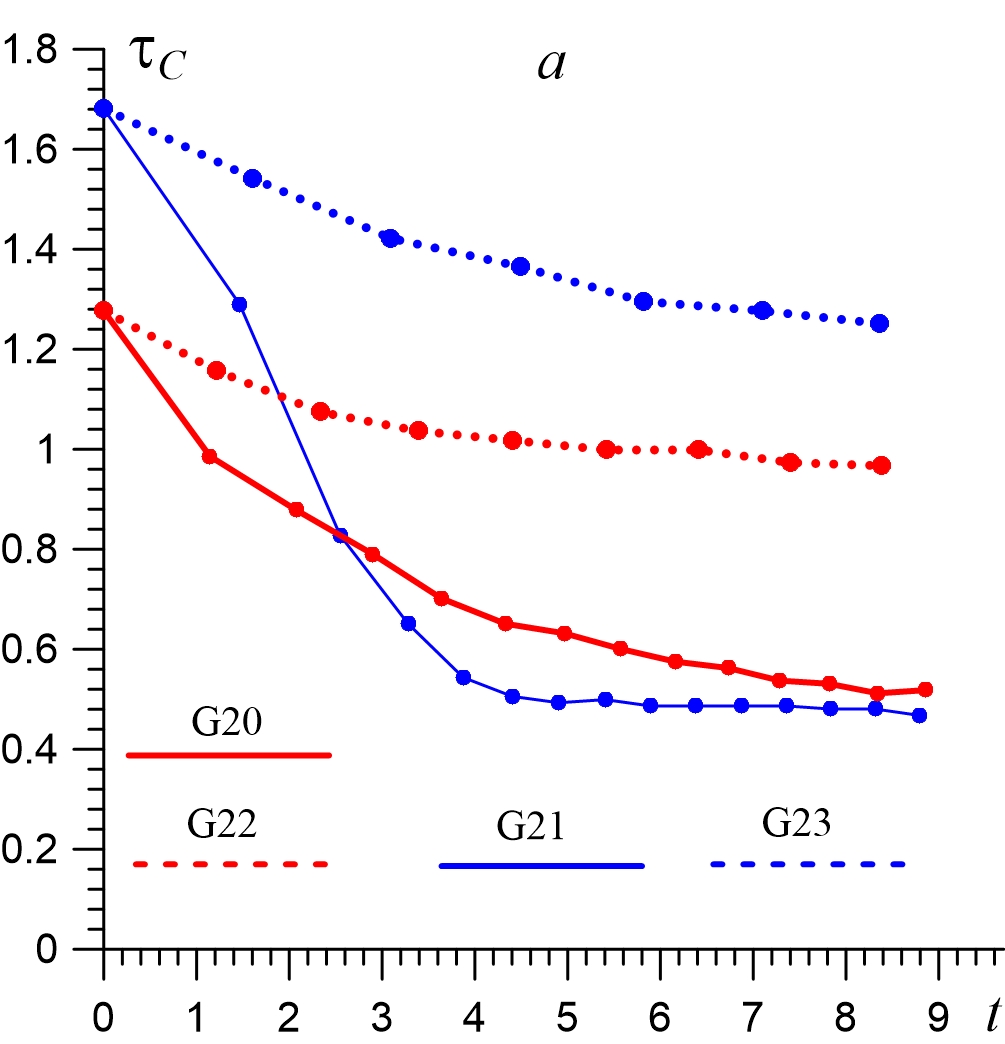} 
\caption{
Periodicity of the passage of the core through the main disc vs. time in different models.
} \label{fig:tau}
\end{figure}

The passage periodicity of the satellite's core through the disc of the main galaxy is characterized by the time $\tau_C$, which depends on the model parameters and changes during the computational experiment (Figure \ref{fig:tau}). The values of $\tau_C$ at the first impacts depend primarily on the initial parameters of the model $r^{(GSE)}_0$, $v_{rot}^{(GSE)}$, $v_r^{( GSE)}$, $\theta^{(GSE)}$ (See Table \ref{tab:Gaia-parameters}). The general trend is a reduction in the time interval $\tau_C$ after the interaction event due to the loss of total energy, which strongly depends on the gas content in the dwarf galaxy.
The period $\tau_C$ changes slightly after gas loss at times $t \gtrsim 4$ billion years (solid lines in Figure~\ref{fig:tau}). The frequency of the disc crossing by the core is significantly lower in models that are initially without gas (dashed lines).

\begin{figure}[h]
\includegraphics[width=0.49\hsize]{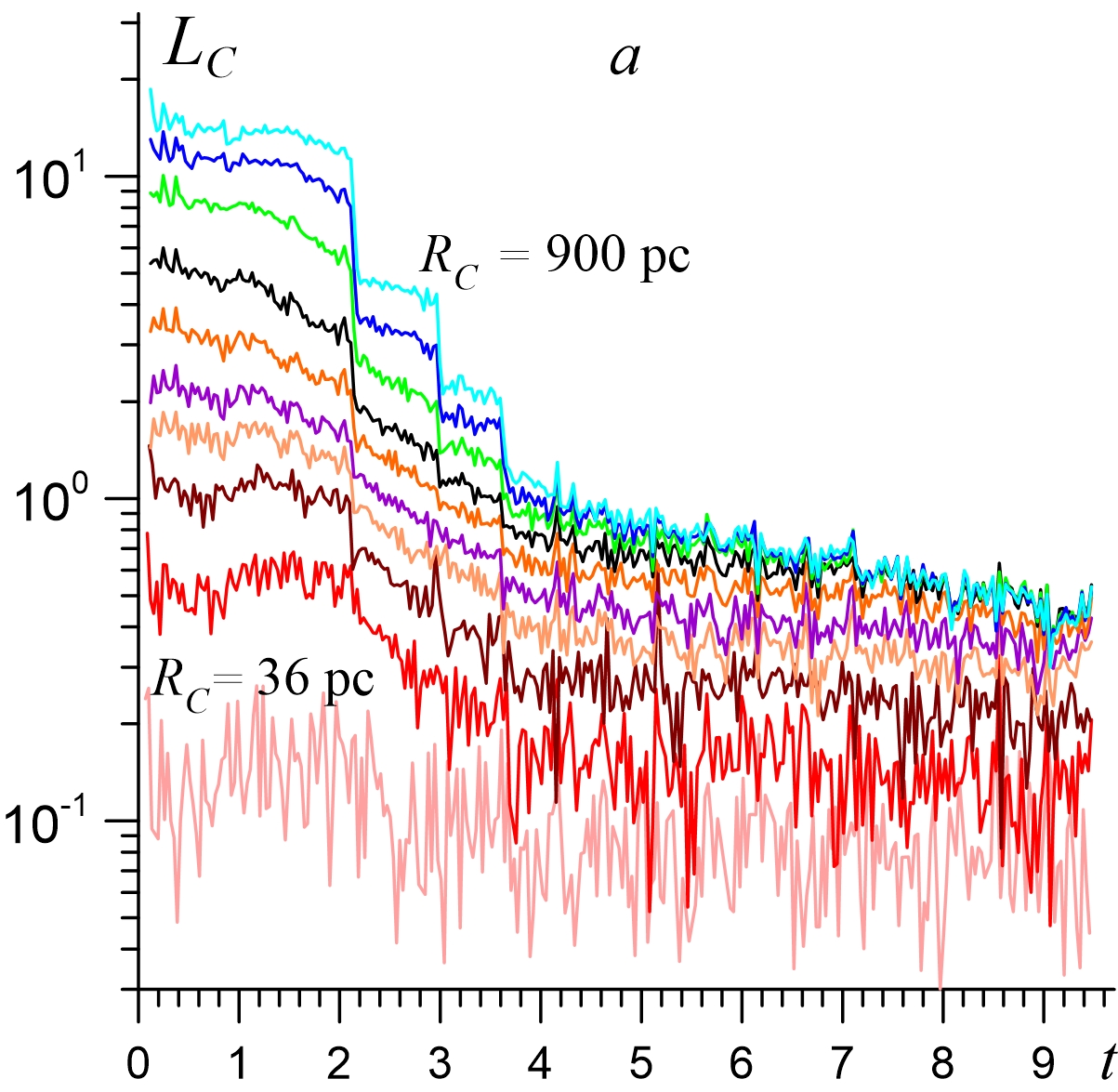} \hskip 0.0\hsize
\includegraphics[width=0.49\hsize]{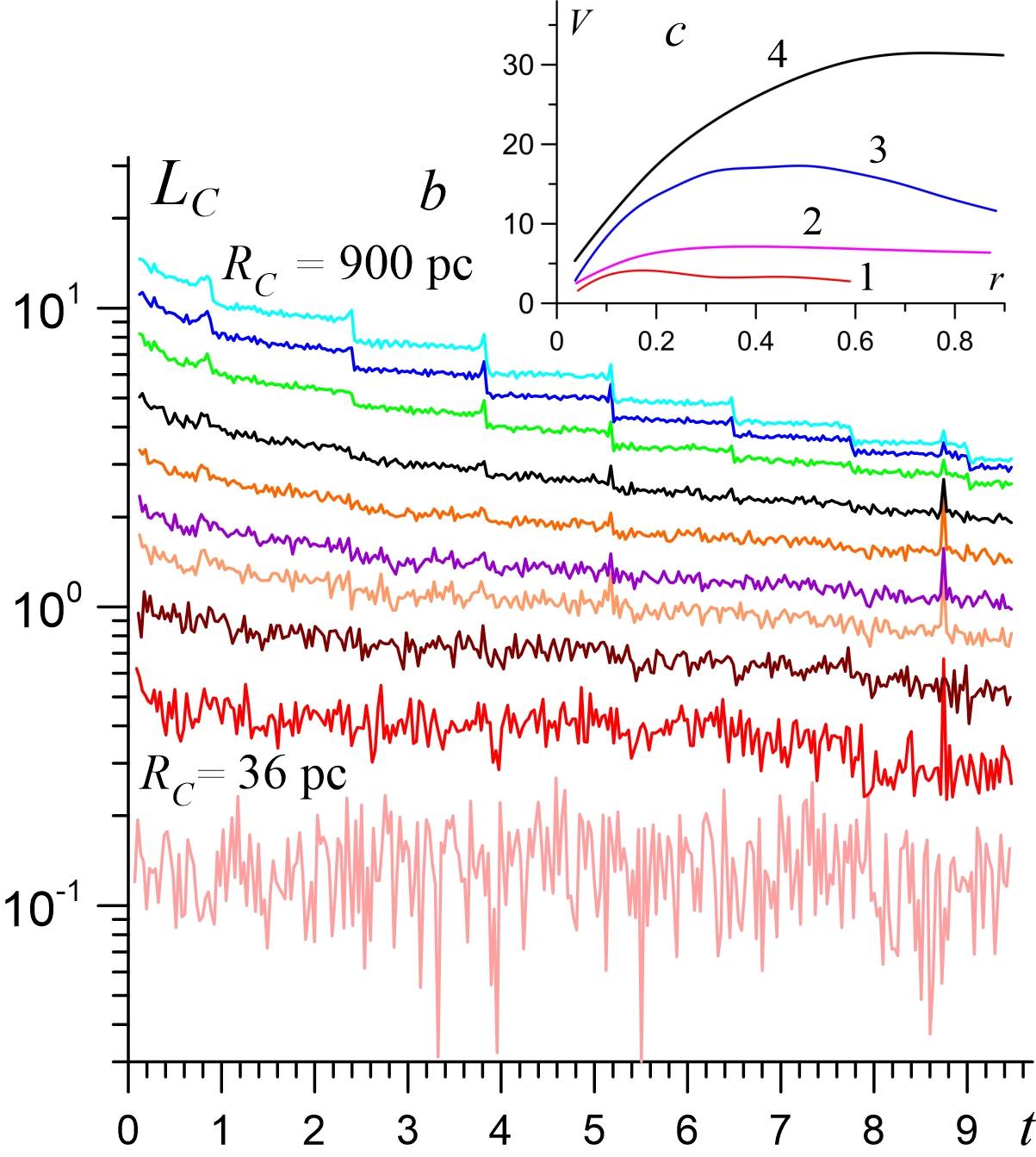}
\caption{
Specific angular momentum of the SC inside different radii vs. time in the models ``G21'' ($a$) and ``G23'' ($b$). The line colors correspond to the conditions in Figure \ref{fig:mass-core-radius}, where the radius $R_C$ varies from 36 pc to 900~pc. $c$ --- Rotation curves of the SC, reconstructed from $L_C(r)$: the model ``G21'' (curve 1 for SC at the end of evolution, curve 2 is after the formation of SC), the model ``G23'' (3 --- SC at the end of evolution, curve 4 after SC formation), $[V]=$km\,sec$^{-1}$.  
} \label{fig:Ls_core_r}
\end{figure}

The process of formation of the elliptical core and peeling of the outer layers is accompanied by a decrease in the specific angular momentum $L_C(t)$ ($[L_C]=$\,kpc\,km\,s$^{-1}$, {
Figure~\ref{fig:Ls_core_r}\,$a$). Thus, the value of $L_C$ decreases due to a decrease in both the size of the core and the internal rotation velocity of the stellar component. After 4 billion years, the core has an almost ellipsoidal shape and a rotation velocity of several kilometers per second, which is approximately an order of magnitude less than the velocity dispersion of stars.  
This result is typical for models with rich initial gas content. Figure \ref{fig:Ls_core_r}\,$b$ shows the evolution of the model ``G23''  without gas for comparison. The rate of angular momentum loss is significantly less, so that a noticeable part of the kinetic energy is associated with the system rotation. The curves in Figure \ref{fig:Ls_core_r}\,$a$ clearly show that angular momentum is lost as a result of the first four passages of the SC through the main galaxy. Then the gas is almost completely swept out of the core (See Figure \ref{fig:Msgdm_core_r-Gaia20-21}\,$b$) and the loss of angular momentum occurs slowly.
 }

\begin{figure}[h]
\includegraphics[width=0.7\hsize]{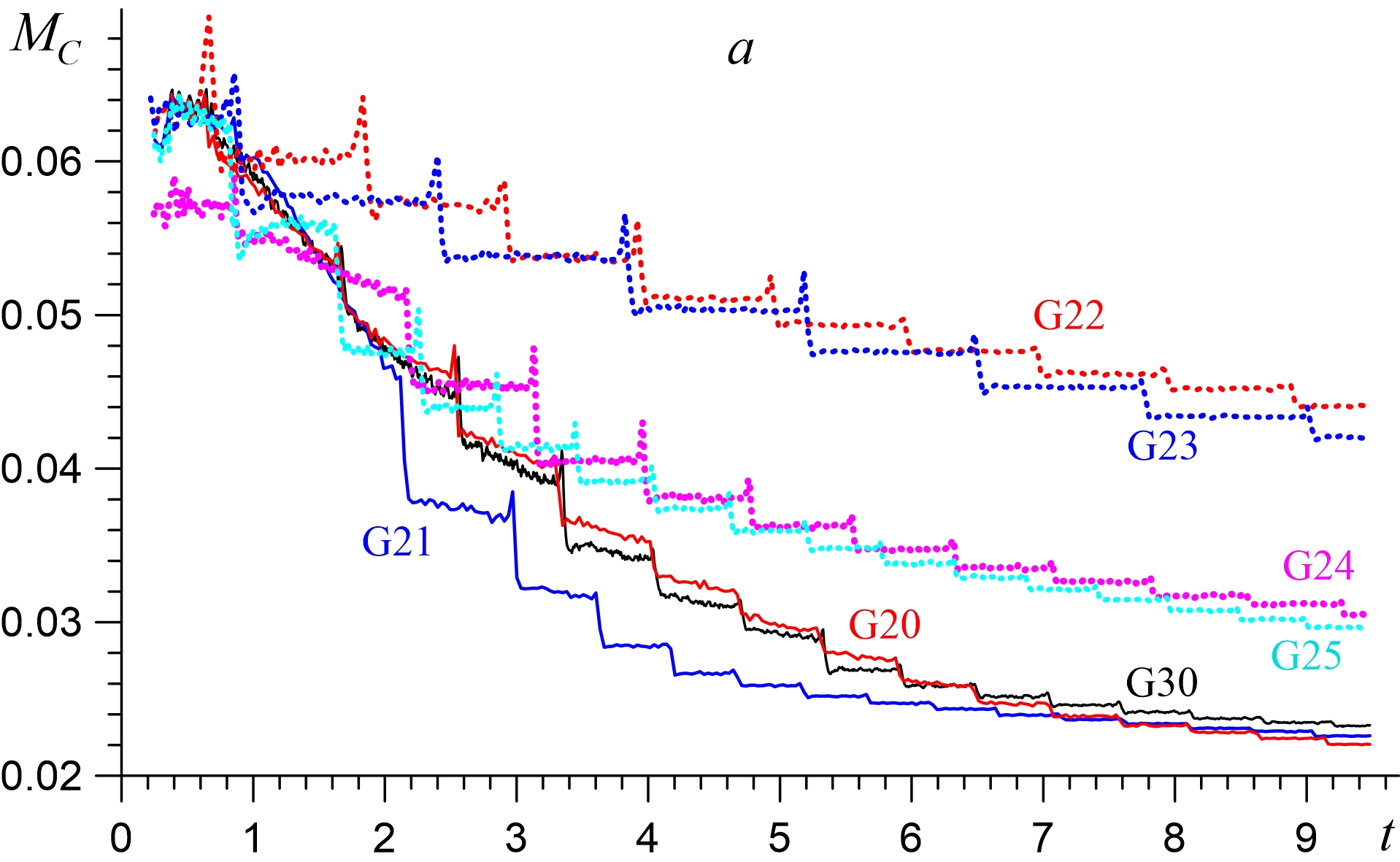} \hskip 0.002\hsize
\includegraphics[width=0.29\hsize]{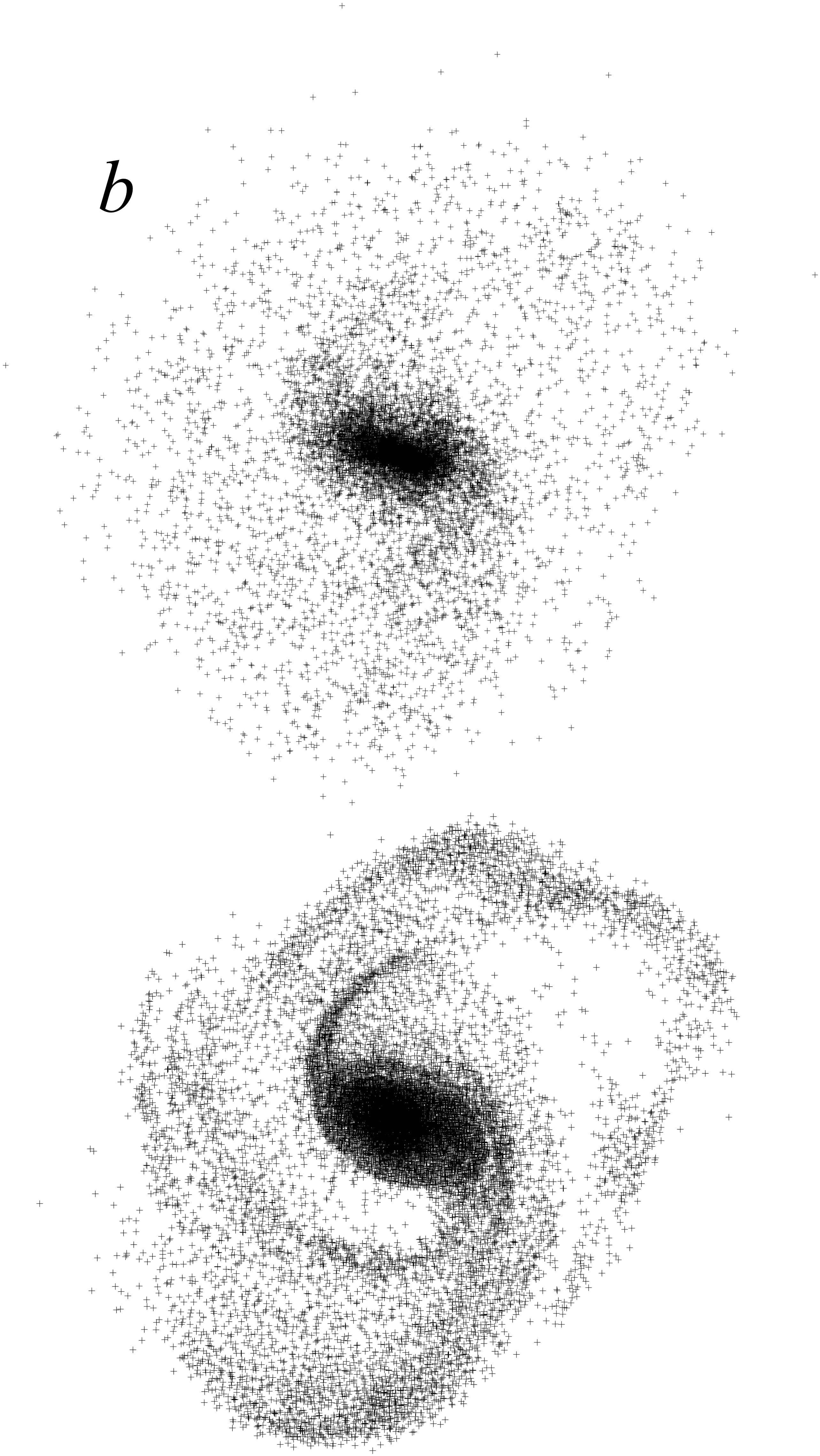}
\caption{$a$ --- Dependence of the total stellar core mass on time in different models. $b$ --- Stellar (top) and gas (bottom) discs of the satellite before their first passage through the MW (model ``G21'').
} \label{fig:total-mass-time}
\end{figure}

Figure \ref{fig:total-mass-time}\,$a$ summarizes the behavior of the total gravitationally bound core mass in different models. { All curves characterize the mass within 900 pc over 9 billion years.} We see characteristic stepwise decreases in mass when crossing the MW disc. Mass loss from the core outside the main galaxy disc is small, except in models with gas, when there is a noticeable decrease in the stellar mass in the SC at high altitudes above the main disc. This effect is clearly visible in the model ``G21'' at $t< 3$ billion years and in the model ``G20'' until approximately 5 billion years. This difference is due to the fact that the SC in the ``G20'' retains gas longer than the model ``G21'' (See Figure \ref{fig:total-mass-time}$a$). The sequence of three curves $M_C(t)$ for models ``G21'' (a lot of gas), ``G24'' (gas halved), ``G23'' (without gas) clearly shows the influence of the gas component on the formation compact stars system. The model ``G30'' is an analogue of the ``G20'' except for the initial orientation of the satellite's disc, which is perpendicular to the disc of the main galaxy. This has little effect on the dynamics of tidal disruption/stripping of the satellite  (See the black line in Figure \ref{fig:total-mass-time}$a$).

A feature of the studied models of minor merging is the presence of the gas and stellar discs in the satellite with a sufficiently powerful stellar bar before the destruction of such a small galaxy begins. Figure \ref{fig:total-mass-time}$b$ shows the distributions of stellar and gas components in the model ``G21'' before the first impact, which are typical for all models considered here. Analysis of the trajectories of all particles leads to the conclusion that the SC is formed primarily from the material of the central region of the stellar bar. The complex rotational motions of the bar particles are destroyed by tidal interactions, and the resulting compact elliptical system retains only a weak relict rotation. The typical ratio of the minor to the major axes are close to 0.85 at the end of evolution.

{ A significant result is that the formed stellar cores of the transitional cE/UCD type are long-lived objects under conditions of periodic flights through a massive large galaxy and can exist for cosmological time scales. }

\begin{figure}[h]
\includegraphics[width=0.49\hsize]{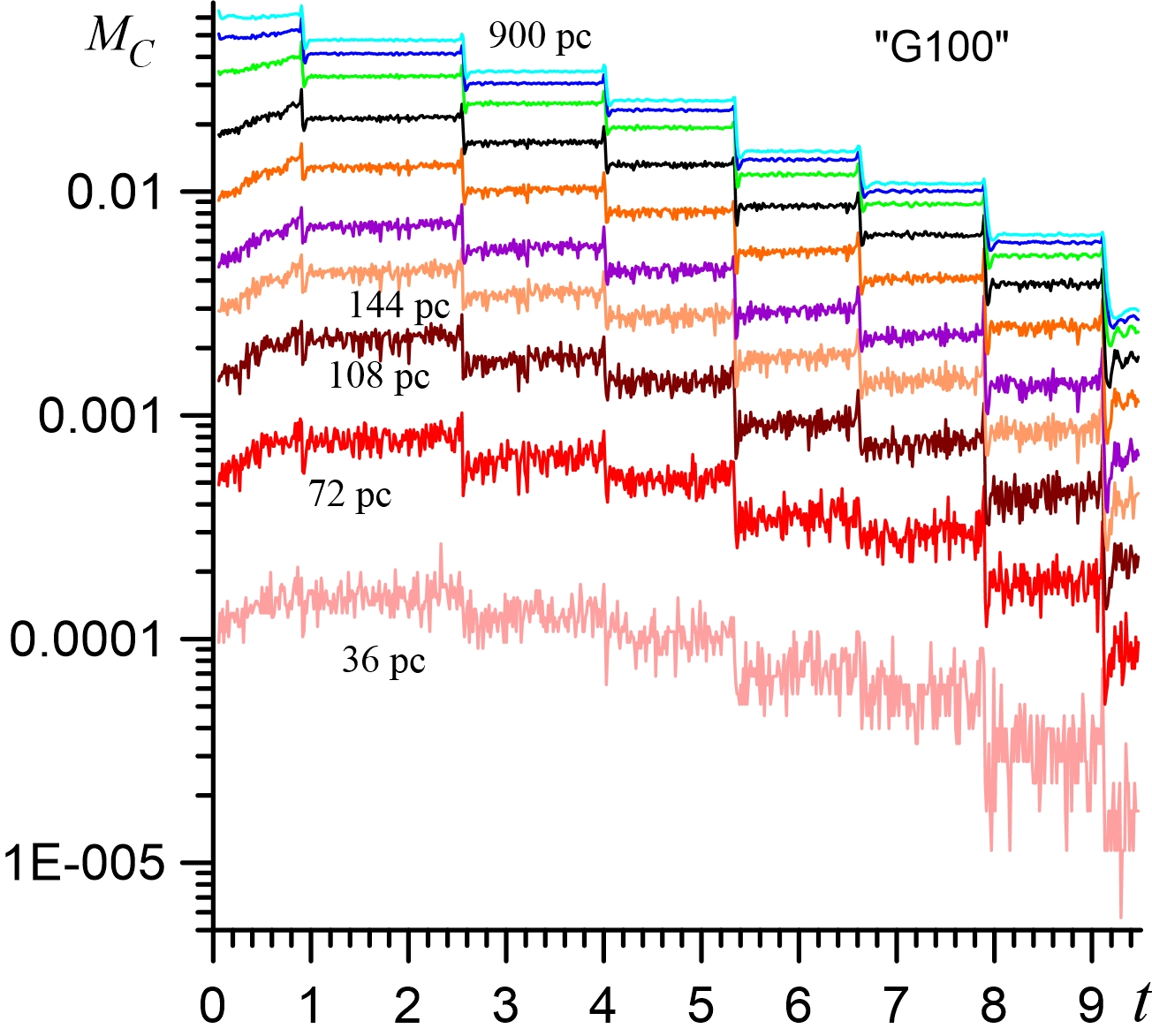} \hskip 0.002\hsize \includegraphics[width=0.49\hsize]{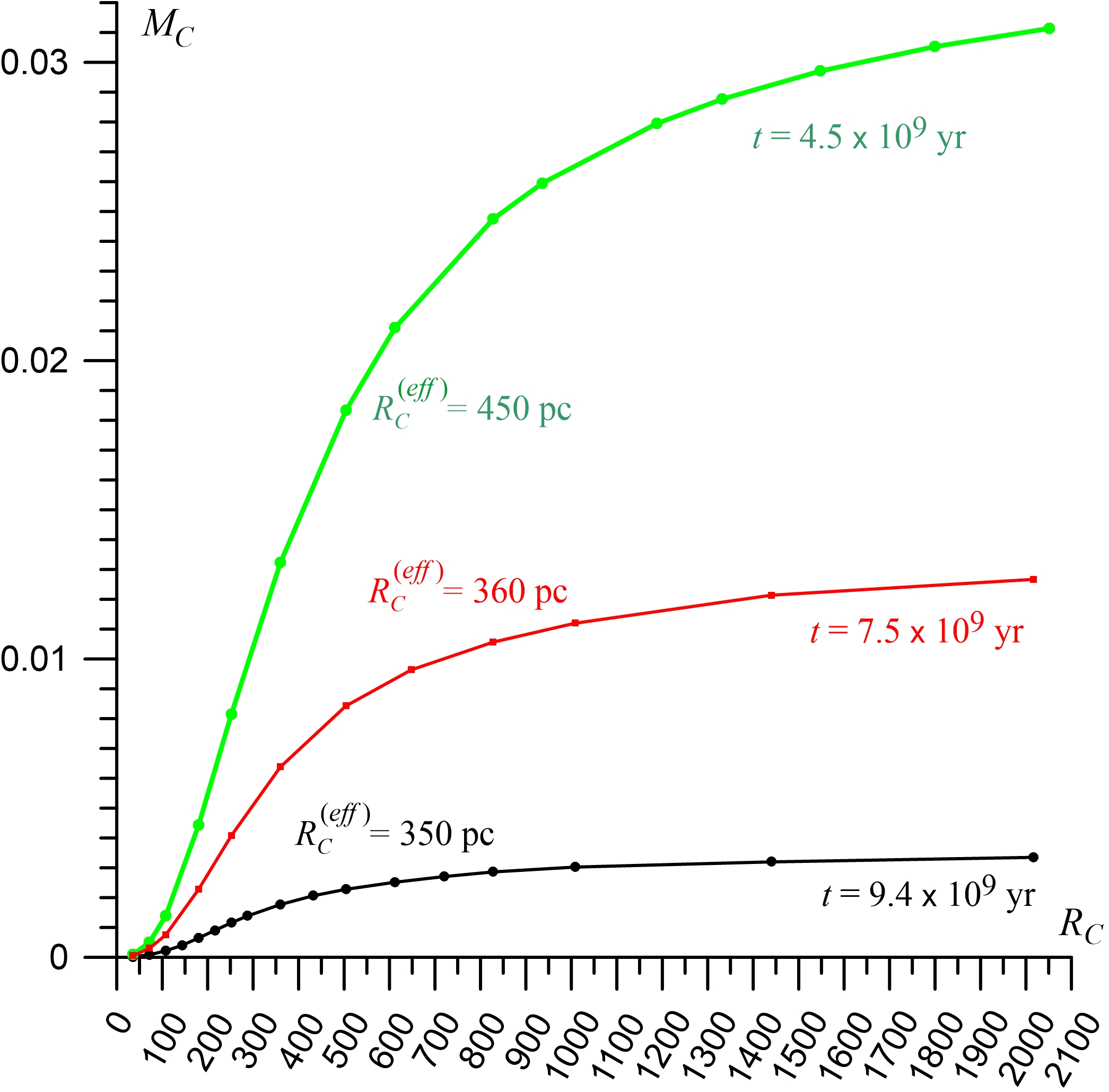} 
\caption{$a$ --- As in Figure \ref{fig:mass-core-radius} for the model ``G100''. $b$ --- Mass of stars inside the sphere with radius $R_C$ at different times (model ``G100'', $[R_C]=$\,pc).
} \label{fig:mass_G100-new}
\end{figure}

{
The final mass of the SC is determined by the initial mass of the satellite and its radial initial stellar density profile. This issue requires some further consideration.
Models ``G20'' -- ``G25'', ``G30'' (See Figures \ref{fig:mass-core-radius} --- \ref{fig:total-mass-time}) contain massive concentrated nucleus. 
Mass density in the central region is low in models ``G46'', ``G49'', ``G50'', ``G100''.  }
{ Figure \ref{fig:mass_G100-new} shows the differences in mass loss rates between the models ``G100'' and ``G23'' (See Figure \ref{fig:Ms_core_r-Gaia23}) due to the lack of an initial concentrated core in model ``G100''. The decrease in mass $M_C$ inside a sphere of fixed radius $R_C$ is several percent at the periphery and does not affect the core region of the model ``G23'' at long times. The model  ``G100'' is distinguished by the absence of an initial concentrated core, which results in a strong stripping of the entire stellar component at all radii throughout the entire evolution (Figure~\ref{fig:mass_G100-new}\,$a$).
Figure~\ref{fig:mass_G100-new}\,$b$ shows the radial dependences of the mass inside a sphere with radius $R_C$, as in the corresponding Figures \ref{fig:mass-core-radius}\,$b$ and \ref{fig:Ms_core_r-Gaia23}\,$b$. Such a SC is too loose compared to the observed UCDs. If the mass $M_C \simeq 3.2\cdot 10^8\,M_\odot$ with an effective radius $R_C^{(ef\!f)} = 450$\,pc at time $4.5\cdot 10^9$\,yr allows us to attribute object to cE type, then the SC mass decreases greatly after 3 billion years with an almost constant effective radius. Thus, the model ``G100'' is an example of the complete destruction of an object of type cE. SC evolves in a similar way in the models ``G49'' and ``G50''. 
 }

\begin{figure}[h]
\includegraphics[width=0.329\hsize]{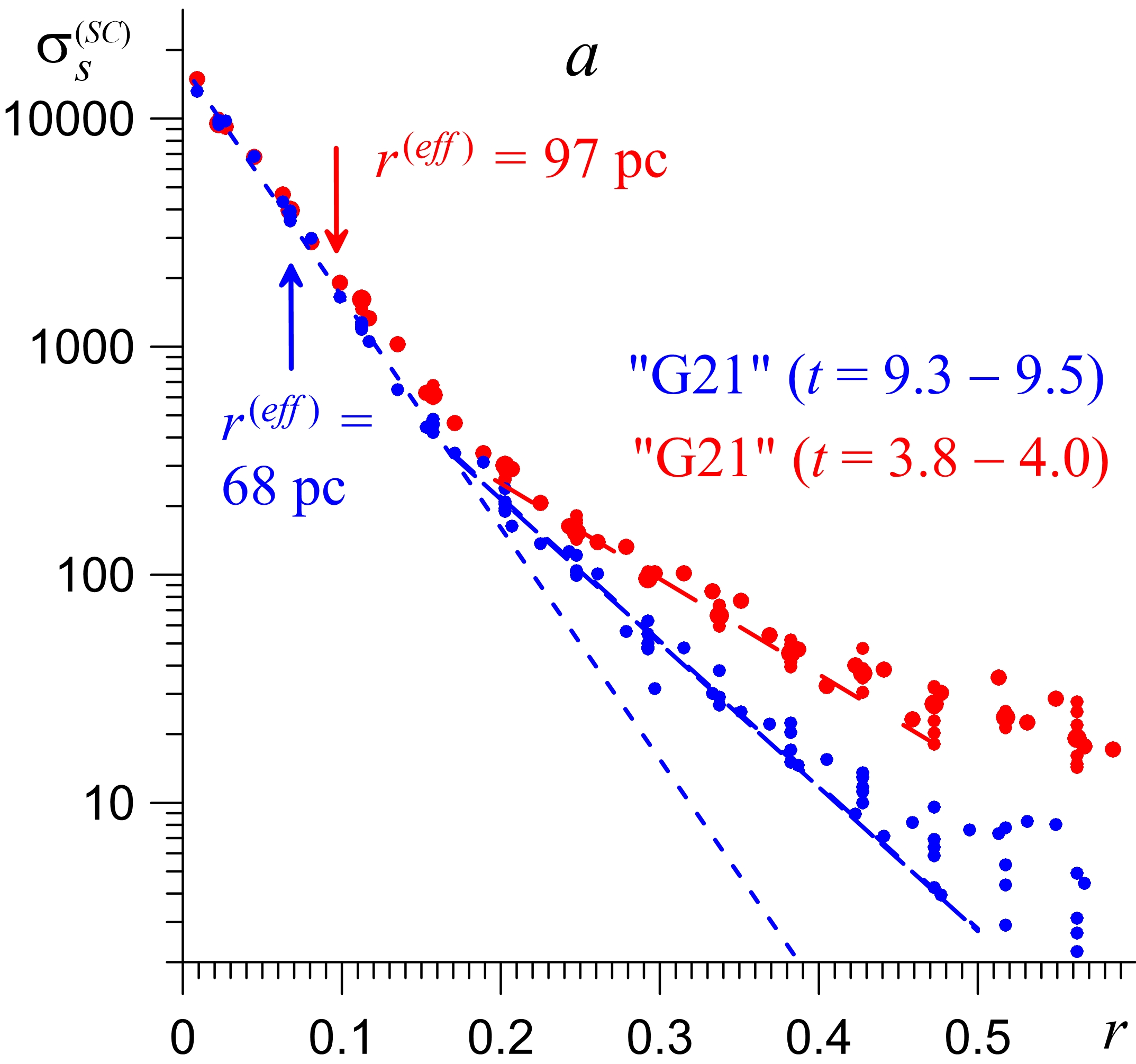}  \includegraphics[width=0.329\hsize]{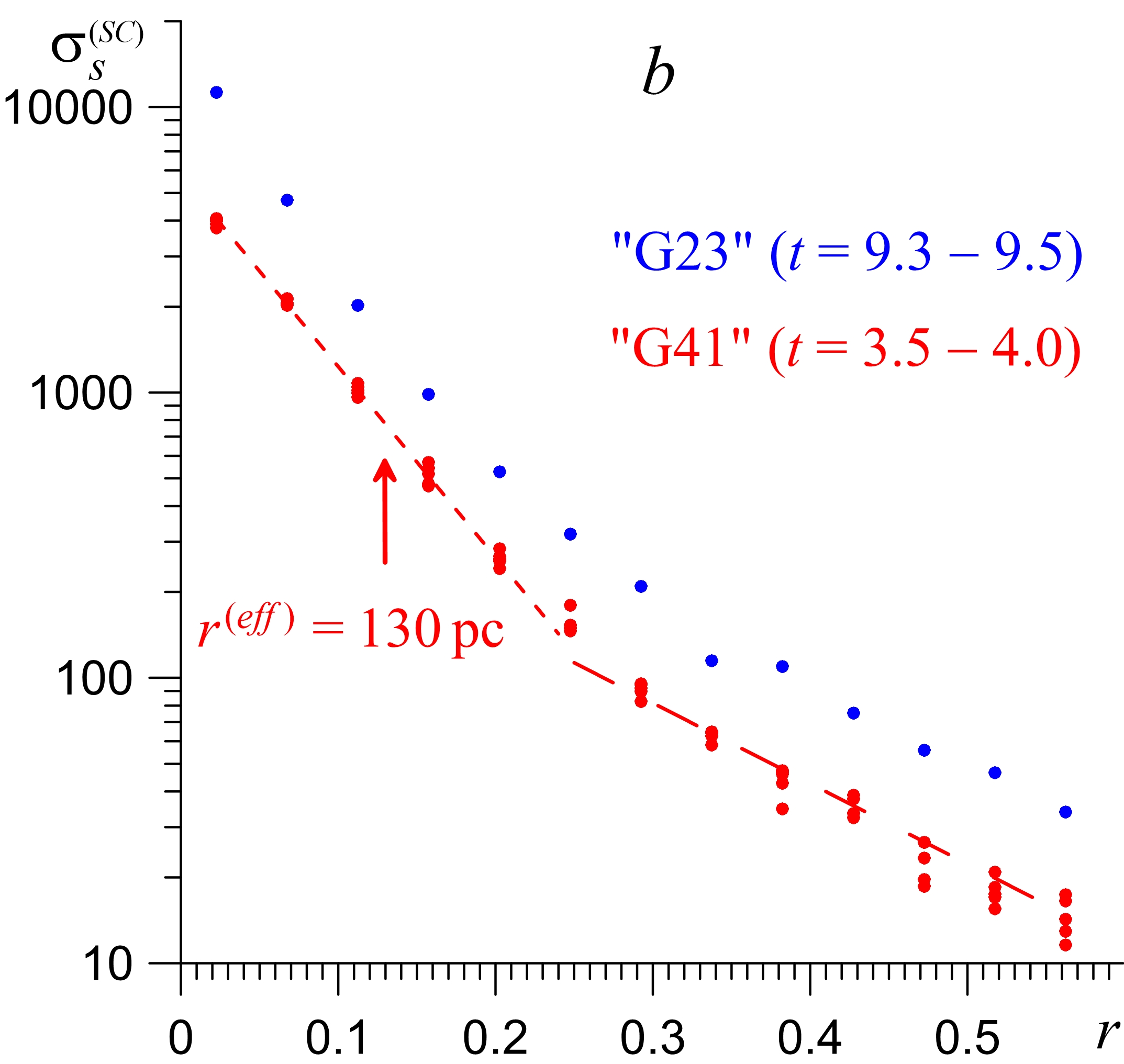} 
\includegraphics[width=0.329\hsize]{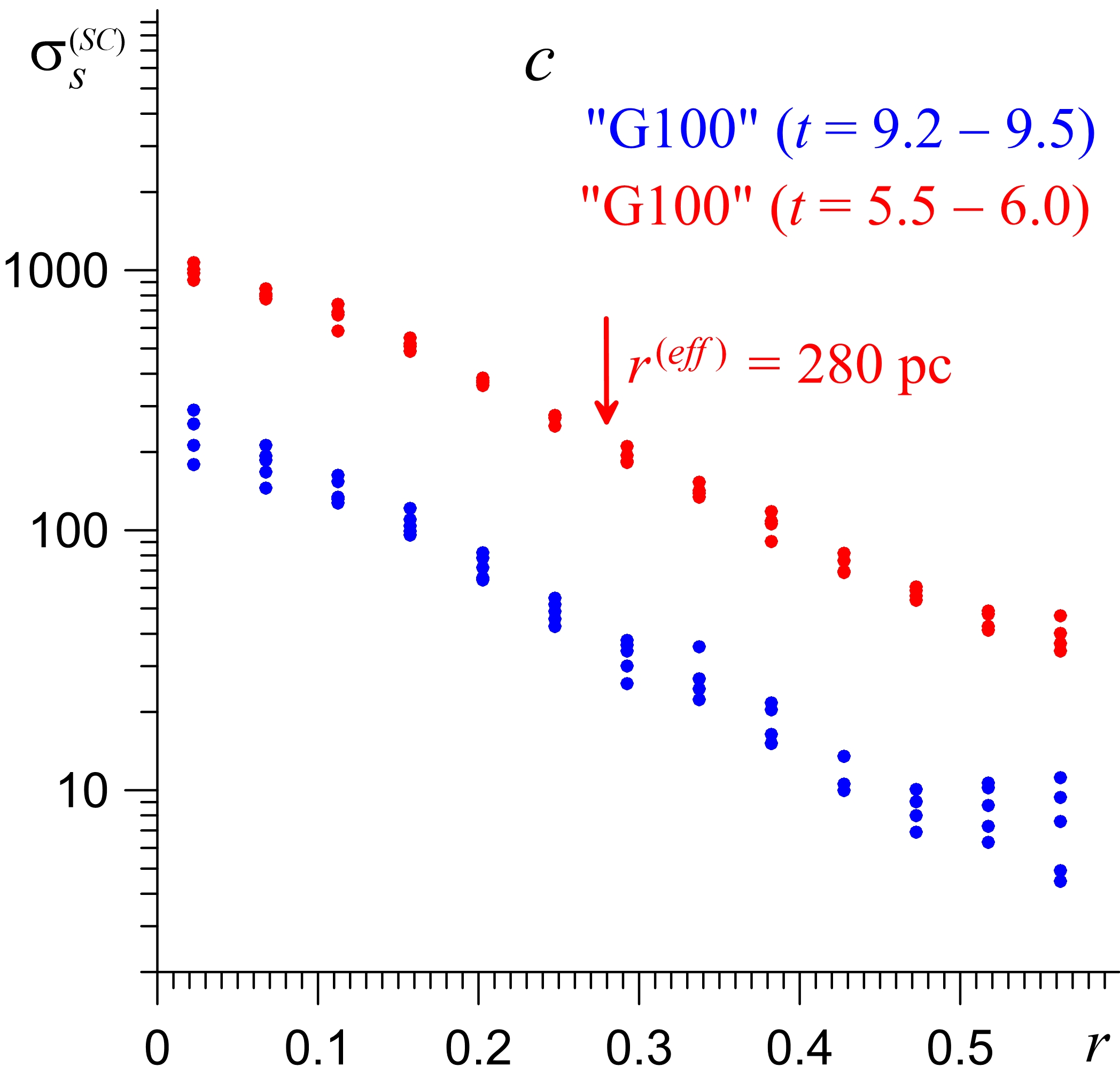}
\caption{Radial profiles of surface density in SCs for different models (circles of different colors). The lines correspond to exponential distributions.
} \label{fig:sigma-GSE-differ}
\end{figure}

{
Let us discuss the radial distributions of surface density of stars ($\sigma_s^{(SC)}(r)$) in SCs as corresponding to the observed characteristics along the line of sight. Figure \ref{fig:sigma-GSE-differ} shows such dependences on the radius $r$ for some models. 
For example, profiles $\sigma_s^{(SC)}(r)$ for the model ``G21'' are shown in Figure \ref{fig:sigma-GSE-differ}\,$a$. Red color shows density distributions after the 4th crossing of the MW. We present several dependences at different times from the interval $t=3.8 - 4$ billion years to show local changes in quantities. Half of the SC's mass for surface density $\sigma_s^{(SC)}(r)$ is inside the effective radius $r^{(ef\!f)} = 97$ pc. The value of $r^{(ef\!f)}$ decreases to 68 pc due to the stripping of the outer layers after the 15th crossing of the host disc. The inner part of the surface density profile ($<100$ pc) remains almost unchanged. Two characteristic zones are distinguished. Both are well approximated by exponential laws with different radial scales. Figure \ref{fig:sigma-GSE-differ}\,$b$ shows the model ``G23'', as an analogue of the ``G21'' without gas, and the model ``G41'', which forms  satellite core with a mass less than 1.5 times compared to ``G21''. 

Finally, Figure \ref{fig:sigma-GSE-differ}\,$c$ contains profiles for  model without an initial concentrated core in a dwarf galaxy. Four passes through the main disc reduce the SC's mass  to $2.28\cdot 10^8\,M_\odot$ with an effective radius $r^{(ef\!f)} = 280$\,pc. The stars content in such an object decreases rapidly. The seventh crossing of the host galaxy leaves approximately $5.5\cdot 10^7\,M_\odot$ in SC at $r^{(ef\!f)} = 383$\,pc, which is not typical for UCD-type. The model ``G49'' with gas without the initial concentrated nucleus leaves even less mass in the SC with a large effective radius. We limit ourselves to constructing the radial distributions of the surface stellar mass density  $\sigma_s^{(SC)}(r)$ in Figure \ref{fig:sigma-GSE-differ}. The transition to surface brightness profiles can be made taking into account the observed mass-to-light ratios $({\cal M}/{\cal L})$, which are $({\cal M}/{\cal L})_B = 3\, (M_\odot/L_\odot)$ and $({\cal M}/{\cal L})_V = 2.5\, (M_\odot/L_\odot)$ (PEGASE.2 models) for the objects under consideration.
 }

\section{Discussion}


The results of modeling the merger of a dwarf barred galaxy with a large Milky Way galaxy are described in detail here. We focus on the evolution of the satellite's dense core, which transforms into a small, compact elliptical star system as its outer layers are stripped away and angular momentum is lost as it repeatedly passes through the host galaxy's disc. Our analysis of the conditions for the formation of compact and ultra-compact elliptical galaxies as a result of minor merging is not comprehensive due to the very large number of free parameters, the variation of which leads to significantly different classes of objects. Our efforts are aimed primarily at studying the influence of the initial gas on the efficiency of stripping the outer (less dense) layers of the forming compact core.

We have considered two extreme classes of dwarf rotating galaxies. One contains a gas-rich disc with a mass comparable to that of stars. The second group of models does not contain gas, as in the case of dwarf lenticular galaxies  (dS0s), in which gas does not have a noticeable effect on the dynamics of the stellar component even in the presence of some star formation \cite{Ge-etal-2021dwarf-S0}. { The baryonic mass of the dwarf galaxy model lies in the range $(1-2)\cdot 10^9\,M_\odot$, which is significantly less than the disc mass of the host galaxy.} The maximum circular velocity is $75-85$ km\,s$^{-1}$ with the optical radius for the stellar component $R_{opt}=2.5-3$\,kpc. 

The loss of a significant portion of the mass of a disc galaxy in the field of a massive elliptical galaxy at the center of rich cluster was demonstrated in the work \cite{Chilingarian-etal-2009cE-Nbody}, where 75 percent of the mass was stripped by tidal forces. Our work is aimed at studying threshing in the field of a large spiral galaxy at large eccentricities of the satellite's orbit. Numerous passages of the dwarf through the disc of MW type galaxy are accompanied by the destruction of its structure and the loss of more than half of the stellar mass, which goes towards the construction of a stellar halo around the main galaxy. The density distribution in such a stellar halo depends significantly on the galactic latitude of arrival of the satellite and is not centrally symmetric.

The formation problem of the galaxy vertical structure and stellar halo through the destruction of the Gaia-Sausage-Enceladus has been actively studied in recent years in connection with the outstanding data of the GAIA project \cite{Anders-etal-2019gaia, Torrealba-etal-2019dwarf-satellite-Gaia, Vieira-etal-2023gaia, Iorio-Belokurov-2019Gaia, Vincenzo-etal-2019SGE}. The spatial picture of the dynamics of all gaseous and stellar components is complex. Therefore, the traditional approach to studying the consequences of merging is based on the properties of phase spaces, for example, energy, angular momentum, action, velocity components and some other parameters 
\cite{Belokurov-etal-2006Sagittarius-Siblings, Cunningham-etal-2022Chemical-Abundances, Feuillet-etal-2020Gaia-Enceladus-Sausage, Limberg-etal-2022Reconstructing-GSE, Amarante-etal-2022Simulated-Chemodynamical-GSE, Khoperskov-etal-2023stellar-halo}. Analysis of the phase distributions of spatial and kinematic characteristics for a large array of Gaia stars makes it possible to identify the history of interactions and mergings in the Milky Way. This method is applicable to data from the results of our minor merges simulations and such studies are in our immediate plans.

Features of the satellite's trajectory significantly influence the merging process, and we are considering the most severe scenario of an almost radial fall, in contrast to a quasi-circular orbits. The lion's share of the satellite's gas is quickly captured by the MW gas disc so that $1-2$ percent of the gas ends up outside the main gas disc and becomes intergalactic medium. This mechanism of enrichment of the intergalactic medium seems interesting for future research. The dwarf's initial dark mass becomes almost entirely part of the main galaxy's halo, with the shape of the additional dark halo generally close to that of the stellar halo.

The presence of gas appears to be an important factor in the birth of compact elliptical dwarfs, promoting the formation of more concentrated core. The physical mechanism for more efficient stripping of the stellar component at an initial high gas content is as follows. The gas makes a noticeable contribution to the gravitational potential of the original dwarf before interaction. During the first and subsequent passages through the disc, the satellite can lose from 10 to 30 percent of the gas, which leads to a decrease in the depth of the gravitational well of the satellite's core. As a result, collisionless particles (stars and DMs) with large velocity dispersion are more easily stripped from the outer layers of the core, forming the stellar halo of the main galaxy.

The low content of dark matter in the core and around the core by the end of evolution (9.5 billion years) is determined, of course, by the initial DM profile in a dwarf galaxy, when the halo mass does not exceed the stellar component mass within the initial radius $R_{opt}^{ (Sat)}$ and the radial scale of the dark halo is larger than the disc scale. The capture of dark matter by the stellar core hardly occurs in our numerical models, since even the stars of the main galaxy do not enrich the core of the satellite. The presence of dark matter in cEs depends on the initial DM radial density scale, and as it decreases, the fraction of dark matter in the core should increase. The paper \cite{Bekki-etal-2003UCD-Fornax} notes that the presence of a central cusp in a massive dark halo makes the formation of compact cores difficult. There are very different ratios of baryonic and dark matter among the observed zoo of cEs. For example, the galaxy PGC 029388 within $r^{(ef\!f)}$ consists of 90 percent DM and is classified as a transition type dSph-UDG with $r^{(ef\!f)} = 1$\,kpc \cite{Afanasiev-etal-2023KDG64}. Compact and ultra-compact elliptical galaxies can exist over cosmological times despite strong tidal interactions. The lifetime of compact stellar cores in our models over several billion years is quite consistent with observational data and other numerical simulations \cite{Bassino-etal-1994Cannibalized-Dwarf-modeling, Akins-etal-2021satellite-MW, García-Bethencourt-etal-2023Milky-Way-simulation, Crain-etal-2023Hydrodynamical-Simulations, Naidu-etal-2021modeling-GSE}.

A wide variety of baryonic to dark matter ratios occur among observed cE galaxies. For example, the galaxy PGC 029388 inside $r^{(ef\!f)}$ consists of DM of 90 percent and is classified as the transition-type dSph-UDG with $r^{(ef\!f)} = 1$ \,kpc \cite{Afanasiev-etal-2023KDG64}.
Compact elliptical galaxies can exist over cosmological times despite strong tidal interactions. The lifetime of compact stellar cores in our models is quite consistent with observational data and other numerical simulations \cite{Bassino-etal-1994Cannibalized-Dwarf-modeling, Akins-etal-2021satellite-MW, García-Bethencourt-etal-2023Milky-Way-simulation, Crain-etal-2023Hydrodynamical-Simulations, Naidu-etal-2021modeling-GSE}.

We limit ourselves to models of a satellite with a small disc and a mass that produces a high concentration of stars in the central region of the disc. This contributes to the formation of more concentrated elliptical galaxies (cE), which are close to UCDs. If the size of the satellite is increased, for example, to 6 kpc, like that of the Large Magellanic Cloud, while maintaining the baryonic mass of $2\cdot 10^9 M_\odot$, then this leads to a decrease in the mass of the core or the complete destruction of the satellite in 9 billion years. The parameters of simulated dwarf elliptical galaxies approach those of ultracompact dwarfs. The construction of UCDs models for a wide range of properties requires additional study.
We plan to develop this study by varying the initial mass distribution of the stellar and gaseous discs in the satellite.
A separate task is the fall of a dwarf galaxy without a bar. This initial model can easily be created by specifying a hotter stellar disc and/or a massive and concentrated dark halo.
However, it is necessary to perceive that tidal forces are capable of generating a central bar when approaching the main galaxy.

{ The discovery of supermassive black holes at the center of compact dwarfs is an important factor in the dynamics of the entire compact star system, since the black holes mass is noticeable compared to the UCDs \cite{Seth-etal-2014BH-UCD, Ahn-etal-2017BH-UCD, Pechetti-etal-2017cE-BH, Afanasiev-etal-2018BH-UCD, Ahn-etal-2018BH-UCD, Voggel-etal-2019BH-UCD, Ferre-Mateu-etal-2021UCD-AGN, Mayes-etal-2023BH-cE}. The black hole helps maintain the rotation of the surrounding stellar component in the core and further increases the concentration of mass density as matter accretes onto the center. Including supermassive black hole in the model of the original satellite for co-evolution during the stripping process could be significant. However, the numerical model used here does not allow us to qualitatively trace the dynamics on scales less than 1 pc. We must significantly increase computational resources to reliably simulate the formation of cE/UCD with a black hole.
}

{ Another additional stripping mechanism may be associated with extragalactic hot gas. Ram pressure can effectively sweep gas out of the galaxy. Beautiful examples are JellyFish galaxies \cite{Ramatsoku-etal-2019jellyfish, Zasov-etal-2020jellyfish}, from which gas comes out in a very long stream due to ram pressure. Gas sweeping by ram pressure from the destroyed satellite changes the total gravitational potential. We showed in this work that this improves the efficiency of stripping the stellar component as well, as the gravitational attraction of the outer layers towards the center is weakened. Simulating minor merging to form compact star systems within hot, rarefied intergalactic gas can improve our understanding of the physics of these objects. 
}

If we look at the GSE event in our Galaxy 10 billion years ago, the results obtained provide certain restrictions on the initial properties of the satellite in the MW. The presence of an increased concentration of stars in the center of a dwarf spiral galaxy before the merger begins ensures the birth of a core with cE/UCD parameters. The system of small satellites in our Milky Way does not contain ultracompact or compact objects, which places restrictions on the hypothetical Gaia-Sausage-Enceladus. The GSE does not appear to have had massive stellar core or bulge in the disc, since we do not see a suitable candidate for a compact or ultra-compact object as a remnant of the GSE.

\section{Conclusions}

We studied the possibility of the formation of compact elliptical galaxies (cE/UCD) as a result of a minor merger of a dwarf disc galaxy and large Milky Way-type spiral galaxy.
Models of the satellite before interaction include stellar and gas components embedded in a dark halo. The maximum circular velocity of the satellite is typical for dwarf galaxies in the range of $75-85$ km\,s$^{-1}$ with an optical radius of about 3 kpc.
The model of the satellite before the collision contains a massive stellar bar and spiral pattern.
The evolution of the interaction is simulated over a period of 9.5 billion years. Each pericentric approach of two objects occurs in the mode of crossing the stellar disc of the main galaxy by the satellite. This ensures the most severe destruction of the dwarf. The ratio of the total masses of two interacting galaxies in the models is 1/46 -- 1/38 within the double optical radius of the stellar discs before the interaction begins. This ratio is 1/30 -- 1/60 for baryonic matter.
The main attention is paid to the features of the formation of a long-lived compact core of the satellite, close to the characteristics of the transitional UCD/cE in mass and size.

The influence of gas on the process of loss of stellar mass by a satellite has been studied in detail. The loss of gas when crossing a large disc galaxy noticeably increases the stripping rate of the stellar component as well. The result of the studied minor merging is transitional cE/UCD galaxies with the following properties.

\noindent --- Approximately half of the initial mass of stars can be used to build a compact stellar core. This fraction decreases with increasing mass of gas in the satellite. The gas component is an important factor contributing to the stripping of the stellar and dark components when passing through the disc of the main galaxy. A gas-rich dwarf produces a compact core with 2 times less mass and size compared to the same satellite without gas. The final gas content in a compact object is negligible.

\noindent --- The formation of a quasi-stationary compact core takes $4-9$ billion years, depending on the gas content and the initial angle of incidence of the satellite on the disc. The fall of a gas-rich dwarf disc galaxy from a region of low galactic latitudes can create cE/UCD galaxy in $4-5$ billion years from just 4 or 5 disc crossings. The fall of a satellite from high galactic latitudes $\theta^{(GSM)}$ extends the time of formation of a compact elliptical galaxy.

\noindent --- All models give strong stripping of the dark halo of the satellite. As a result, the DM mass decreases by more than an order of magnitude and does not have a noticeable effect on the dynamics of the satellite’s core. Thus, the formed cE/UCD models do not contain significant dark mass at the end of the numerical simulations.

\noindent --- The constructed models of compact elliptical galaxies differ slightly from the spherical shape and the ratio of the axes gives the morphological type E0 --- E2 depending on the initial gas content in the satellite.

\noindent --- { The compact/ultra-compact cores in our numerical models have a mass of $(1 - 5)\cdot 10^{8}M_\odot$ and an effective radius of $r^{(ef\!f)} = 60-200$\,pc, and are long-lived. } Such objects continue to persist almost unchanged over dozens of orbital periods despite further numerous crossings of the disc of the main galaxy. The core stellar mass loss per revolution $\Delta{M_C}/M_C$ can initially exceed $\sim 30$ percent in gas-rich models and decreases greatly with each new crossing of the main galaxy. The result of one crossing of the disc by the core in the modern era after 9 billion years of evolution is $\Delta{M_C}/M_C \simeq 1-2$ percent in the initially gas-rich satellite and $3-6$ percent in the gas-free satellite.

{\noindent --- The rotation of stars in stripped nuclei is determined by the initial gas content in the satellite in the considered models. A very gas-rich satellite produces cE/UCD-type objects with almost no rotation. Models without initial gas give noticeable internal rotation of the compact object. It should be noted that different mechanisms for stripping the nucleus of a dwarf galaxy can result in different rates of angular momentum loss. For example, our consideration of interaction with a large spiral galaxy may be very different from merging in a cluster with a cD galaxy. 
 }


\vspace{6pt} 




\authorcontributions{
Conceptualization, A.V.K.; methodology, S.S.K.; software, S.S.K.; validation, D.S.S., S.S.K.; formal analysis, A.V.K.; investigation, D.S.S., S.S.K.; data curation, S.S.K. and D.S.S.; writing---original draft preparation, A.V.K.; writing---review and editing, A.V.K.; visualization, S.S.K. and D.S.S.; supervision, A.V.K.; project administration, A.V.K.; funding acquisition, A.V.K. All authors have read and agreed to the published version of the manuscript.
 }

\funding{
This work supported by the Russian Science Foundation~(grant no. 23-71-00016,  \\ https://rscf.ru/project/23-71-00016/). The research also relied on the shared research facilities of the HPC computing resources at Lomonosov Moscow State University.
 }

\acknowledgments{
We thank reviewers for their valuable comments and suggestions.
 }




\abbreviations{Abbreviations}{
The following abbreviations are used in this manuscript:\\

\noindent 
\begin{tabular}{@{}ll}
MW & Milky Way\\
GSE & Gaia-Sausage-Enceladus \\
GSEC & Gaia-Sausage-Enceladus Core \\
DM & Dark Matter\\
SPH & Smoothed-Particle Hydrodynamics \\
UCDs & Ultracompact Dwarf Galaxies \\
UDGs & Ultra-diffuse galaxies \\ 
 GCs & Globular clusters \\
 NSC & Nuclear Star Cluster \\
SC & Satellite Core \\
\end{tabular}
}


\begin{adjustwidth}{-\extralength}{0cm}

\reftitle{References}


\bibliography{Bib}

%


\PublishersNote{}
\end{adjustwidth}
\end{document}